\documentclass[aps,prb,amsmath,amssymb,footinbib,showpacs,twocolumn]{revtex4-2}
\usepackage{amsmath}
\usepackage{amssymb}
\usepackage{amsthm}
\usepackage{setspace}
\usepackage{graphicx}
\usepackage{braket}
\usepackage{mathrsfs}
\usepackage{float}
\usepackage[colorlinks = true,linkcolor = blue,urlcolor  = blue,citecolor = blue,anchorcolor = blue]{hyperref}
\usepackage[utf8]{inputenc}
\usepackage[english]{babel}
\usepackage{bm}

\begin{document}

\title{Quantum interference of pseudospin-1 fermions}

\author{Adesh Singh}
\affiliation{School of Physical Sciences, Indian Institute of Technology Mandi, Mandi 175005, India}
\author{G. Sharma}
\affiliation{School of Physical Sciences, Indian Institute of Technology Mandi, Mandi 175005, India}

\begin{abstract}
Quantum interference is studied in a three-band model of pseudospin-one fermions in the $\alpha-\mathcal{T}_3$ lattice. We derive a general formula for magnetoconductivity that predicts a rich crossover between weak localization (WL) and weak antilocalization (WAL) in various scenarios. Recovering the known results for graphene ($\alpha=0$), we remarkably discover that WAL is notably enhanced when one deviates slightly from the graphene lattice, i.e. when $\alpha>0$, even though Berry's phase is no longer $\pi$. This is attributed to the presence of multiple Cooperon channels. Upon further increasing $\alpha$, a crossover to WL occurs that is maximal for the case of the Dice lattice ($\alpha=1$). Our work distinctly underscores the role of non-trivial band topology in the localization properties of electrons confined to the two-dimensional $\alpha-\mathcal{T}_3$ lattice.
\end{abstract}

\maketitle
The interference of waves is so fundamental to physics that it unites various branches, including but not limited to optics, acoustics, quantum mechanics, solids, and cold atoms. In solids, if the disorder is sufficiently high, wave interference can lead to a complete suppression of electronic transport. This phenomenon is known as Anderson localization (AL)~\cite{anderson1958absence}. A precursor to AL, weak localization (WL) refers to the negative quantum correction to the Drude conductivity due to the interference of electron waves~\cite{altshuler1980magnetoresistance,bergmann1984weak}. In WL theory, the deviation to the conductivity is expanded in terms of the parameter $\lambda_F/l$ ($\lambda_F$ and $l$ being the Fermi wavelength and the mean free path, respectively). Since disorder is inevitable in nature, WL is the standard method to measure the phase coherence length as well as determine the relevant processes responsible for electron scattering~\cite{chakravarty1986weak,beenakker1991quantum}. Coupling the electrons to the magnetic field introduces a finite phase difference between the interfering waves, and hence, magnetoconductivity is a critical tool in the study of WL. In contrast to conventional WL, the presence of spin-orbit coupling can lead to phase shift via spin precession resulting in destructive interference of electron waves, thereby enhancing the conductivity. This phenomenon, termed weak antilocalization~\cite{hikami1980spin} (WAL), also occurs in Dirac and Weyl materials where pseudospin replaces the actual spin~\cite{suzuura2002crossover,khveshchenko2006electron,mccann2006weak,gorbachev2007weak,wu2007weak,gorbachev2007weak,tikhonenko2008weak,tikhonenko2008weak,tkachov2011weak,lu2011competition,lu2013intervalley,lu2014finite,fu2019quantum}.

The Dirac and Weyl equations that originated in particle physics now describe the low-energy physics of materials such as graphene, Weyl semimetals, and Van der Waal structures, leading to their resurgence in condensed matter physics~\cite{neto2009electronic,sarma2011electronic,armitage2018weyl}. The band topology of these materials makes them of high interest, which leads to peculiar properties. For instance, the presence of $\pi$ Berry phase leads to the WAL effect in graphene~\cite{suzuura2002crossover,khveshchenko2006electron,mccann2006weak}. Recent experimental breakthroughs in Van der Waal heterostructures, such as the discovery of twisted bilayer graphene exhibiting flat bands, have further intensified research in this arena~\cite{cao2018unconventional}. 

\begin{figure}
    \centering
    \includegraphics[width=0.99\columnwidth]{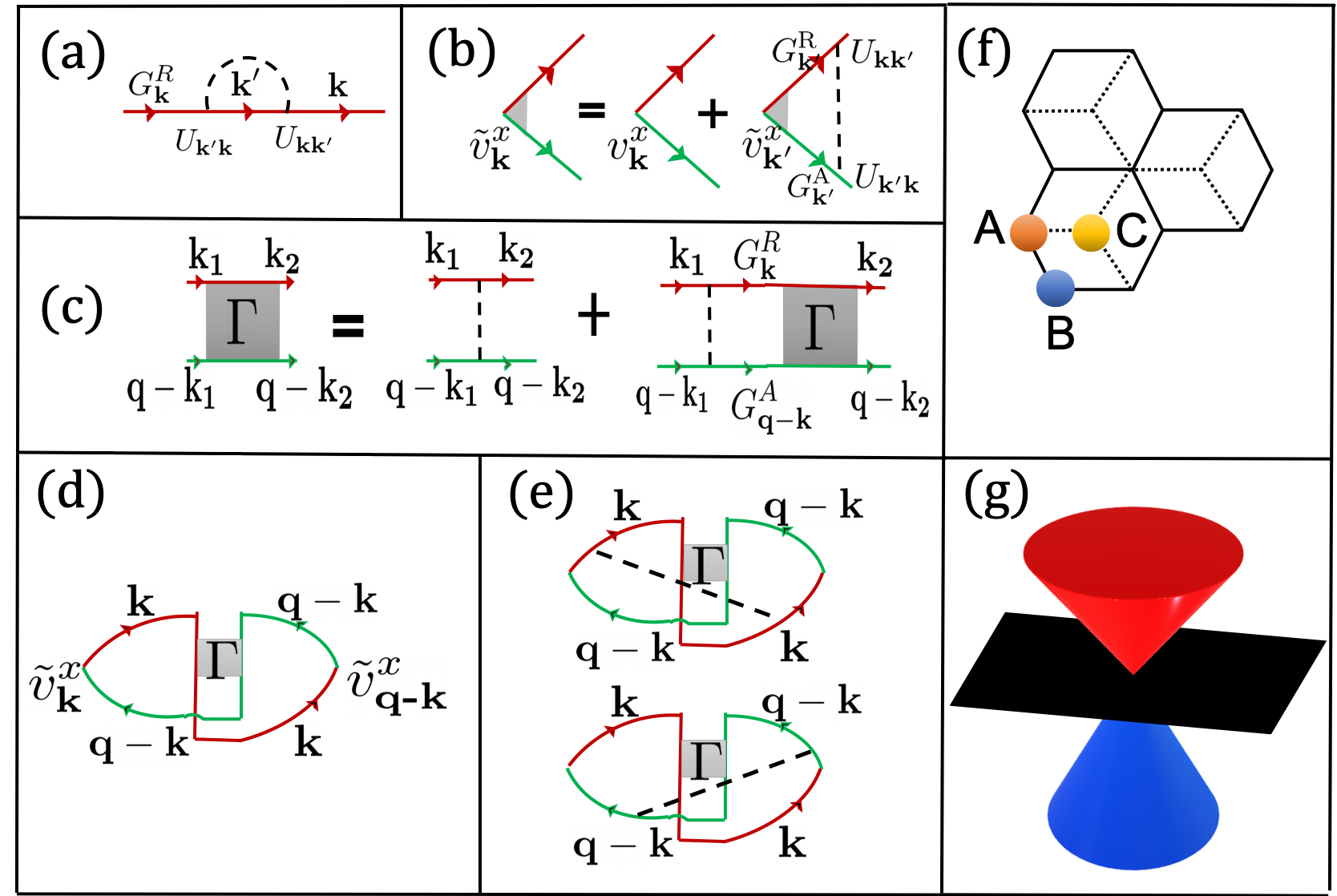}
    \caption{(a) The retarded Green's function. (b) Vertex correction to the velocity. (c) The bethe-Salpeter equation for the vertex. (d) Bare, and (e) two dressed Hikami boxes for calculation of conductivity. (f) The $\alpha-\mathcal{T}_3$ lattice. The hopping between sublattices $A$ and $B$, and $A$ and $C$ is $t$ and $\alpha t$ respectively  (g) The bandstructure of the $\alpha-\mathcal{T}_3$ lattice model with a flat-band in the middle intersecting the Dirac cones.}
    \label{fig:feyn-diag}
\end{figure}

Almost all the quantum interference studies in these materials so far have been typically based on a two-band model that mimics Dirac and Weyl physics. Specifically, it was pointed out in the two-band Dirac fermion that a crossover from weak antilocalization to weak localization occurs as the $\pi$ Berry phase reduces to zero with the inclusion of Dirac mass~\cite{lu2011competition}. The problem of quantum interference effects on localization properties in multi-band models is largely unexplored despite the prevalence of such systems in experiments. For instance, the $\alpha-\mathcal{T}_3$ lattice model~\cite{raoux2014dia} that synthesizes the Dirac and flat-band physics in a single model comprises a hexagonal lattice with atoms situated at the vertices of the hexagons and their centers, thus describing a three-band system of pseudospin-1 fermions (Fig.~\ref{fig:feyn-diag} (f). By varying the hopping parameter between two sublattices, one interpolates between graphene ($\alpha=0$) and the dice lattice ($\alpha=1$). The $\alpha-\mathcal{T}_3$ model can be realized in trilayers of cubic lattices, Hg$_{1-x}$Cd$_x$Te quantum wells and cold-atom systems~\cite{wang2011nearly, rizzi2006phase, malcolm2015magneto, serret2002vortex}.

In this Letter, we solve the problem of quantum interference in pseudospin-1 fermions and examine the localization properties of electrons in the $\alpha-\mathcal{T}_3$ lattice, deriving a general formula for magnetoconductivity that predicts a rich crossover between localization and antilocalization in various scenarios. We first recover the known results for graphene ($\alpha=0$). For the case of only elastic impurities, we remarkably discover that weak antilocalization is notably enhanced when one deviates slightly from the graphene lattice (i.e. when $\alpha>0$), even though Berry's phase is no longer $\pi$. We attribute this behavior to the presence of two Cooperon channels.
Upon further increasing $\alpha$, we crossover to weak localization that is maximal for the case of Dice lattice ($\alpha=1$). Since the band structure is independent of $\alpha$, our model distinctly highlights the role of non-trivial band topology in the localization properties of electrons confined to the two-dimensional $\alpha-\mathcal{T}_3$ lattice.

We consider the following model of pseudospin-1 fermions in the  $\alpha-\mathcal{T}_3$ lattice~\cite{raoux2014dia}:
\begin{equation}
 H^\mu(\mathbf{k})=
 \begin{pmatrix}
 0& a f_{\mu}(\mathbf{k}) & 0\\
 a f_{\mu}^{*}(\mathbf{k}) & 0&  b f_{\mu}(\mathbf{k})\\
 0 &  b f^{*}_{\mu}{ (\mathbf{k})} &0
\end{pmatrix}
\label{Eq_Hamiltonian}
\end{equation}
where $f_{\mu}(\mathbf{k})=\mu \hbar v_F(k_{x}-ik_{y})$, $\mu=\pm1$ is the valley index, $v_{F}$ is the velocity parameter, $\psi=\tan^{1}(\alpha)$ with $a=\cos\psi$, $b=\sin \psi$. The energy dispersion is given by $\epsilon_\mathbf{k} = 0$, $+\hbar v_F k$, and $-\hbar v_F k$, corresponding to a flat zero-energy band intersecting the linearly dispersing Weyl cone (Fig.~\ref{fig:feyn-diag} (g)). We focus on the case when the Fermi level intersects the conduction band, in which case the corresponding eigenfunction is given by $\psi_\mathbf{k}(\mathbf{r})= (1/\sqrt{2})[\mu ae^{-i\mu\phi}, 1, \mu be^{i\mu\phi}] e^{i\mathbf{k}\cdot\mathbf{r}}$. 

We consider both elastic and (pseudo)magnetic impurities such that the impurity potential is given by $U(\textbf{r})=U_{0}(\textbf{r})+U_{m}(\textbf{r})$, where $U_{0}(\textbf{r})$ the elastic scattering potential, and $U_{m}(\textbf{r})$ is for the magnetic scattering potential. We assume point-like disorder with $U_{0}(\textbf{r})=\sum_{\textbf{R}_{i}}u^{i}_{0} S_0\delta(\textbf{r}-\textbf{R}_{i})$, 
$U_{m}(\textbf{r})=\sum_{\textbf{R}_{i}}\sum\limits_{\alpha=x,y,z}u^{i}_{\alpha}S_{\alpha}\delta(\textbf{r}-\textbf{R}_{i})$, where we sum over impurity potentials located at random positions $\textbf{R}_{i}$, $S_0\equiv \mathbb{I}_3$, $\mathbf{S}=(S_{x},S_{y},S_{z})$ is the vector of spin-1 matrices, and $u^i_{0,\alpha}$'s are the corresponding impurity potentials~\cite{SI}. While comparing our results to the particular case of graphene ($a\rightarrow 1$), we must note that the spin matrices $S_i$'s do not reduce to the two-component Pauli spin matrices $\sigma_i$'s. For example, when $a\rightarrow 1$, the sublattice $C$ is wholly decoupled, and while the $S_z$ matrix couples with energy $u_z$ to sublattice A and zero energy to sublattice B, different from a $\sigma_z$ impurity where the coupling would be $u_z$ to sublattice A and $-u_z$ to sublattice B. Thus when focusing on graphene with magnetic impurities, we must draw the comparison carefully, although we don't find qualitative differences in the results. 
We neglect the interference of different impurities with each other. In what follows, we assume in-plane isotropy ($u_x=u_y$), although, in Ref.~\cite{SI}, we also present results for the general case.

Quantum interference to conductivity in pseudospin-1 fermions is calculated diagrammatically (see Fig.~\ref{fig:feyn-diag} (a)-(e)). The retarded (R) and advanced (A) Green's functions are: $ G^{R/A}_{\textbf{k}}(\omega)={1}/{(\omega-\epsilon_{\textbf{k}}\pm i{\hbar}/{2\tau})}$, where the scattering rate is given by $\tau^{-1}=\tau_e^{-1}+\tau_z^{-1}+2\tau_x^{-1}$; ${\hbar}{\tau}^{-1}_{e}=  {2\pi N_{F}}{n_{0}u^{2}_{0}}(a^{4}+b^{4}+1)/4$ defines the elastic scattering rate, while ${\hbar}{\tau_{x}^{-1}}= {2\pi N_{F}}{n_{m}u^{2}_{x}}/{2}$, and ${\hbar}{\tau_{z}^{-1}}= {2\pi N_{F}}{n_{m}u^{2}_{z}}(a^4+b^4)/{4}$ define the magnetic scattering rates. Furthermore, $\tau_{e}$ is related to the elastic scattering length $\ell_{e}$ by $\ell_{e}=\sqrt{D\tau_{e}}$, where $D=v_F^2\tau/2$ is the diffusion constant. We similarly define the magnetic scattering lengths $l_x$ and $l_z$ as well.
The vertex correction to the velocity (Fig.~\ref{fig:feyn-diag} (b)) is evaluated by self-consistently solving the following equation: $\tilde{v}_{\mathbf{k}}^{i}=v_{\mathbf{k}}^{i}+\sum_{\mathbf{k}^{\prime}} G_{\mathbf{k}^{\prime}}^{\mathrm{R}} G_{\mathbf{k}^{\prime}}^{\mathrm{A}}\left\langle U_{\mathbf{k}, \mathbf{k}^{\prime}} U_{\mathbf{k}^{\prime}, \mathbf{k}}\right\rangle_\textrm{imp}\tilde{v}_{\mathbf{k}^{\prime}}^{i}$, where $\tilde{v}_\mathbf{k}^i=\eta_\nu v_\mathbf{k}$, and $\eta_\nu$ is evaluated to be~\cite{SI} 
\begin{align}
\eta^{-1}_{v}={1-\left({\alpha_e}/{(a^{4}+b^{4}+1)}+2ab{\alpha_x}\right)}.
\label{Eq_eta_nu}
\end{align}
We recover $\eta_\nu=2$ for graphene\cite{suzuura2002crossover,mccann2006weak}. The net conductivity is evaluated by summing over the contribution of one bare and two dressed Hikami boxes (Fig.~\ref{fig:feyn-diag} (d)-(e))~\cite{SI}:
\begin{align}
\sigma= -\frac{ e^{2}N_{F}\tau^{3}\eta^{2}_{v}v^{2}_{F}}{\hbar^{2}}(1+2\eta_{H})\sum_{\textbf{q}}\Gamma(\textbf{q}),
\label{Eq_conductivity}
\end{align}
where $N_F = {E_{F}}/{2\pi(\hbar v_{F})^2}$ is the density of states,  $\eta_{H}=-({1}/{2})\left(1-\eta^{-1}_{v}\right)$, and the vertex $\Gamma(\mathbf{q})$ is evaluated by solving the Bethe-Salpeter equation $ \Gamma_{\mathbf{k}_{1}, \mathbf{k}_{2}}= \Gamma_{\mathbf{k}_{1}, \mathbf{k}_{2}}^{0}+\sum_{\mathbf{k}}\Gamma_{\mathbf{k}_{1}, \mathbf{k}}^{0} G_{\mathbf{k}}^{R}G_{\mathbf{q}-\mathbf{k}}^{A} \Gamma_{\mathbf{k}, \mathbf{k}_{2}}$, where $\mathbf{q}=\mathbf{k}_1+\mathbf{k}_2$. We recover $\eta_H=-1/4$ for graphene. 
In the limit when $\mathbf{q}\rightarrow 0$, the bare vertex is evaluated to be $\Gamma^{0}_{\mathbf{k_{1},k_{2}}}\equiv \langle U_{\mathbf{k_1},\mathbf{k}_{2}}U_{\mathbf{-k_1},\mathbf{-k_2}}\bigl \rangle_{\mathrm{imp}} = ({\hbar}/{2\pi N_{F}\tau})\sum_{m}\sum_{n}z_{mn}e^{im\phi_{1}}e^{in\phi_{2}}$, where both $m$ and $n$ run between -2 to +2, and $z_{mn}$ are entries of the matrix $z$ given by~\cite{SI} 
\begin{align}
     z= 
    \begin{pmatrix}
    0 & 0 & 0 & 0 & z^{(-22)}\\
     0 & 0 & 0 & z^{(-11)} & 0 \\
      0 & 0 & z^{(00)} & 0 & 0 \\
      0 & z^{(1-1)} & 0 & 0 & 0 \\
      z^{(2-2)} & 0 & 0 & 0 & 0 &
    \end{pmatrix},
\end{align}
and the elements are given by $ z^{(-22)} = \frac{b^4 \alpha _e}{a^4+b^4+1}+\frac{b^4 \alpha _z}{a^4+b^4}$, $
z^{(-11)} = \frac{2 b^2 \alpha _e}{a^4+b^4+1}-2 b^2 \alpha _x $, $z^{(00)} = \frac{\left(2 a^2 b^2+1\right) \alpha _e}{a^4+b^4+1}-\frac{2 a^2 b^2 \alpha _z}{a^4+b^4}-4 a b \alpha _x $, $z^{(1-1)} = \frac{2 a^2 \alpha _e}{a^4+b^4+1}-2 a^2 \alpha _x $, and $z^{(2-2)} = \frac{a^4 \alpha _e}{a^4+b^4+1}+\frac{a^4 \alpha _z}{a^4+b^4}$. 

The Bethe-Salpeter equation (Fig.~\ref{fig:feyn-diag} (c)) is solved using the ansatz $\Gamma_{\mathbf{k_{1},k_{2}}}=({\hbar}/{2\pi N_{F}\tau})\sum_{m}\sum_n=\gamma^{mn}e^{im\phi_{1}}e^{in\phi_{2}}$, where the coefficients $\gamma^{mn}$ of the matrix $\gamma$ are solved by the equation $\gamma=(I-z\Phi)^{-1}z$, and $\Phi^{mn}= \frac{1}{2\pi}\int_{0}^{2\pi}{e^{i(m+n)\phi}}({1+i\tau\mathbf{q}\cdot \mathbf{v}_{F}})^{-1} d\phi$. We evaluate~\cite{SI}
\begin{align}
 &{\gamma^{(-mm)}}=\nonumber\\
 &\frac{2 \prod\limits_{p\neq m} g^{(-pp)}}{\prod\limits_k \left(g^{(-kk)}+Q^2 \left(\sum\limits_l \frac{1}{g^{(-ll)}} + \sum\limits_{q}\frac{1}{g^{(-qq)}g^{(-q-1,q+1)}}\right) \right) } ,  
 \label{Eq_gamma}
\end{align}
where the ``Cooperon gaps" have been introduced $g^{(-ii)} = 2(1-z^{(-ii)})/z^{(-ii)}$, and $Q=qv_F\tau$. Eq.~\ref{Eq_gamma} along with Eq.~\ref{Eq_conductivity} form the main results of this paper. We evaluate the vertex $\Gamma(\mathbf{q})$ retaining Cooperon gaps that result in diverging contributions to the conductivity~\cite{SI}.

The conductivity is evaluated by integrating Eq.~\ref{Eq_conductivity} between $1/l_e$ and $1/l_\phi$~\cite{suzuura2002crossover}. In the presence of a magnetic field, the wavevector $q$ is quantized as $q_n^2 = (n+1/2) 4eB/\hbar$, where $n$ is the Landau level index. Summing over $n$ and subtracting the zero-field conductivity gives us magnetoconductivity $\Delta\sigma(B)$~\cite{bergmann1984weak}. We present magnetoconductivity results focusing on the weak-$B$ regime, i.e., $l_B^2\gg l_e^2$.

\begin{figure}
    \centering
    \includegraphics[width=\columnwidth]{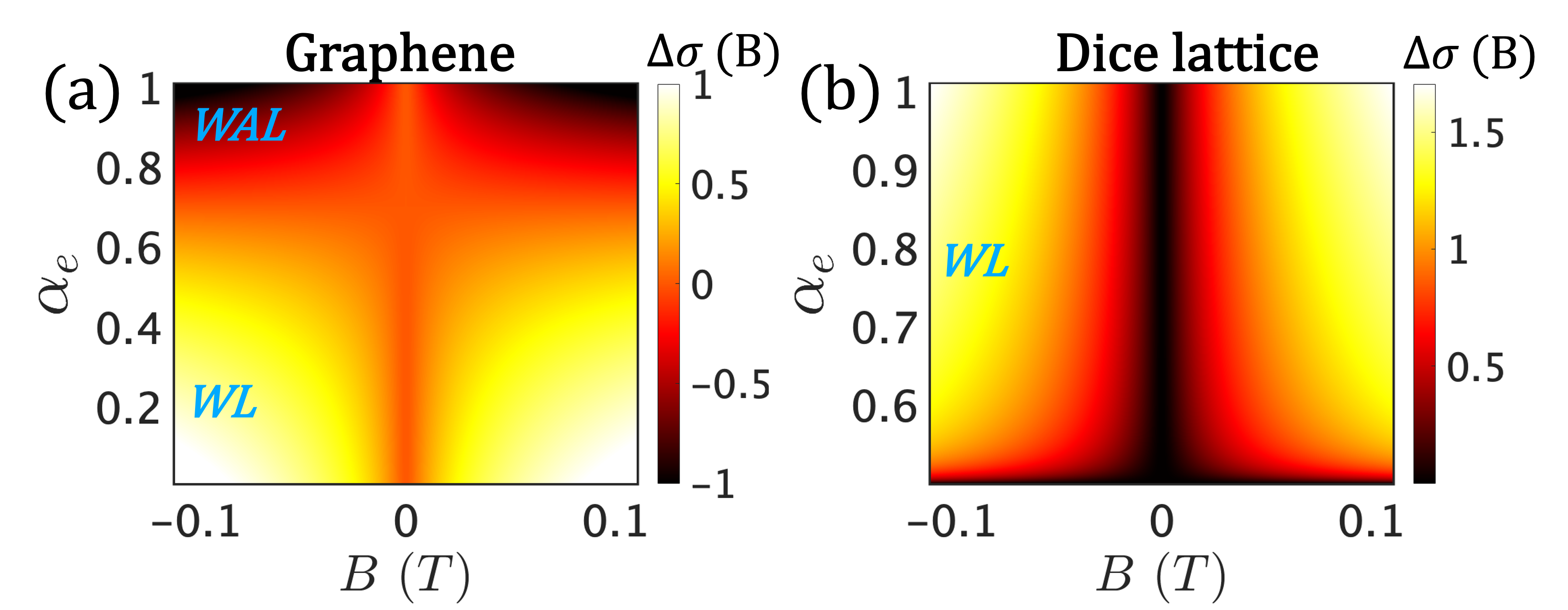}
    \caption{Magnetoconductivity in the units of $e^2/\pi\hbar$. (a) Crossover from weak localization to weak antilocalization in graphene as $\alpha_e$ is increased, i.e., the relative scattering rate of the elastic impurities is increased compared to magnetic impurities. (b) Dice lattice displays only weak localization. We chose $l_\phi=300$ nm and $l_e=1000$ nm.}
    \label{fig:delsigmaB_graphene_dice}
\end{figure}

In the case of graphene ($a=1$), the Cooperon gaps that can vanish are $g^{(1-1)}$, which results in WAL and $g^{(2-2)}$ that results in WL. We evaluate the magnetoconductivity to be~\cite{SI} 
\begin{align}
    &\Delta\sigma(B) =\nonumber\\
    &\frac{e^{2}}{\pi h}\sum_{i=0, 1}\alpha_{i}\left[\Psi\left(\frac{\ell^{2}_{B}}{\ell^{2}_{\phi}}+\frac{\ell^{2}_{B}}{\ell^{2}_{i}}+\frac{1}{2}\right)-\ln\left(\frac{\ell^{2}_{B}}{\ell^{2}_{\phi}}+\frac{\ell^{2}_{B}}{\ell^{2}_{i}}\right)\right],
\end{align}
where $\Psi$ is the digamma function, and
\begin{align}
      &\alpha_{0}= \frac{\eta^{2}_{v}(1+2\eta_{H})}{2\left(1+\frac{1}{g^{(1-1)}}\right)}, \quad \alpha_{1}= -\frac{\eta^{2}_{v}(1+2\eta_{H})}{2\left(\frac{1}{g^{(2-2)}}+\frac{1}{g^{(00)}}+1\right)},& \nonumber\\
     &{\ell_{0}}^{-2} = \frac{g^{(2-2)}}{2\ell^{2}\left(1+\frac{1}{g^{(1-1)}}\right)}, \quad \ell^{-2}_{1} = \frac{g^{(1-1)}}{2\ell^{2}\left(\frac{1}{g^{(2-2)}}+\frac{1}{g^{(00)}}+1\right)},&
\end{align}
where $l^{-2}=l_e^{-2}+l_z^{-2}+2l_x^{-2}$. We observe a crossover from weak localization to weak antilocalization as $\alpha_e$ varies from zero to unity. Magnetic impurities suppress WAL in graphene, as expected~\cite{lu2011competition}. We plot this behavior in Fig.~\ref{fig:delsigmaB_graphene_dice} (a). Note that the other solution of graphene ($a=0$) is similar; the cooperon gaps that vanish are $g^{(-11)}$, which results in WAL and $g^{(-22)}$ that causes WL.
\begin{figure}
    \centering
    \includegraphics[width=\columnwidth]{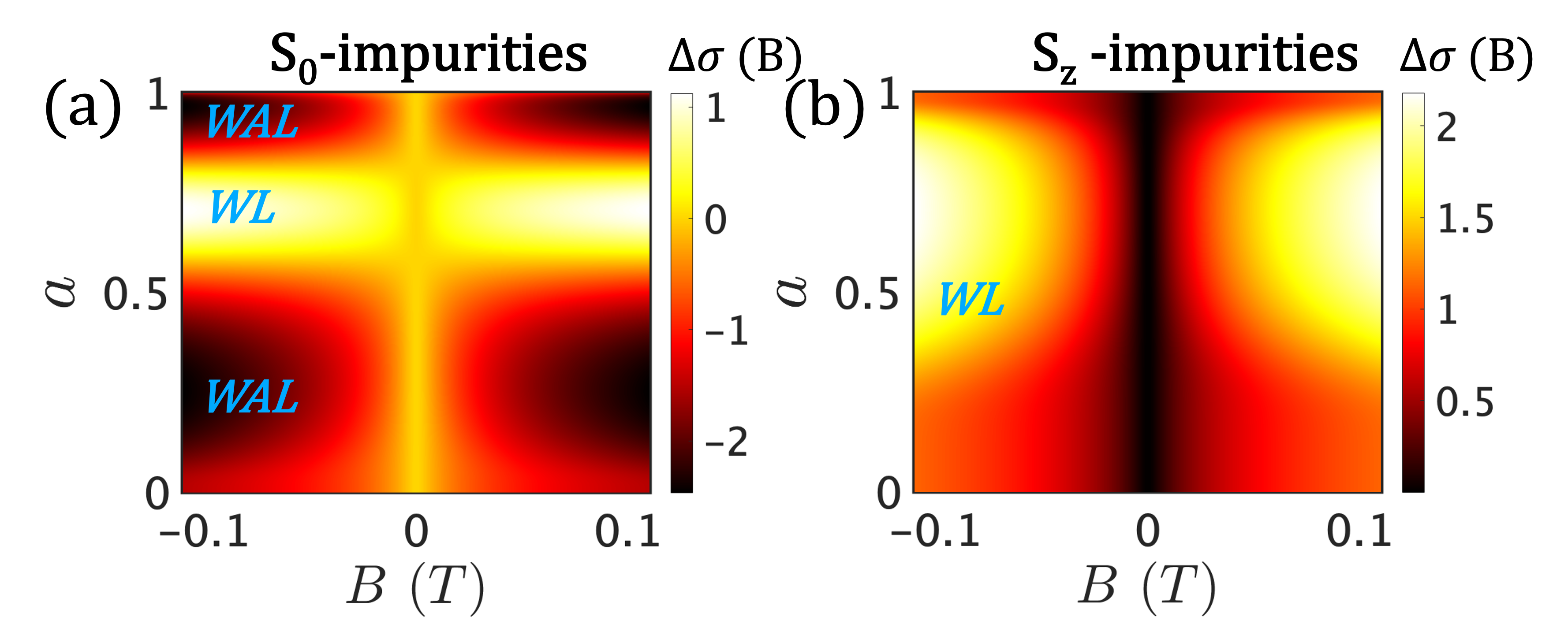}
    \caption{Magnetoconductivity in the units of $e^2/\pi\hbar$. (a) Crossover from WAL ($a=0$, graphene) to WL ($a=2^{-\frac{1}{2}}$, dice) to WAL ($a=1$, graphene) for elastic impurities ($\alpha_e=1$). WAL is maximum slightly away from the graphene lattice, while WL is maximum for the dice lattice. (b) WL for the case of only $S_z$ impurities. We chose $l_\phi=300$ nm and $l_e=l_z=1000$ nm.}
    \label{fig:delsigmaB_elastic-mag-z}
\end{figure}

In the case of dice lattice ($a=2^{-\frac{1}{2}}$), the only vanishing Cooperon gap is $g^{(00)}$, which causes WL. The magnetoconductivity is evaluated to be~\cite{SI}
\begin{align}
    &\Delta\sigma(B) = \nonumber \\ 
    &\frac{e^{2}}{\pi h}\alpha_{0}\left[\Psi\left(\frac{\ell^{2}_{B}}{\ell^{2}_{\phi}}+\frac{\ell^{2}_{B}}{\ell^{2}_{i}}+\frac{1}{2}\right)-\ln\left(\frac{\ell^{2}_{B}}{\ell^{2}_{\phi}}+\frac{\ell^{2}_{B}}{\ell^{2}_{i}}\right)\right],
\end{align}
where
\begin{align}
  \alpha_{0}= \frac{\eta^{2}_{v}(1+2\eta_{H})}{2\left(1+\frac{1}{g^{(-11)}}\right)}, \quad \ell^{-2}_{0} = \frac{g^{(00)}}{2\ell^{2}\left(1+\frac{1}{g^{(-11)}}\right)},    
\end{align}
In contrast to graphene lattice, dice lattice displays only weak localization, as seen in Fig.~\ref{fig:delsigmaB_graphene_dice} (b). This can be understood as the quantized Berry phase changes from $\pi$ to zero as we change the hopping parameter from $a=1$ (graphene) to $a=2^{-\frac{1}{2}}$ (dice lattice). 

In the presence of only elastic impurities ($\alpha_e=1$), the Cooperon gaps that vanish are $g^{(1-1)}$, $g^{(-11)}$, both of are WAL channels, and $g^{(00)}$, which is WL channel. We evaluate the magnetoconductivity to be~\cite{SI} 
\begin{align}
    &\Delta\sigma(B) =\nonumber \\ 
    &\frac{e^{2}}{\pi h}\sum_{i=0,1, 2}\alpha_{i}\left[\Psi\left(\frac{\ell^{2}_{B}}{\ell^{2}_{\phi}}+\frac{\ell^{2}_{B}}{\ell^{2}_{i}}+\frac{1}{2}\right)-\ln\left(\frac{\ell^{2}_{B}}{\ell^{2}_{\phi}}+\frac{\ell^{2}_{B}}{\ell^{2}_{i}}\right)\right],
\end{align}
where
\begin{align}
      &\alpha_{0}= -\frac{\eta^{2}_{v}(1+2\eta_{H})}{2\left(1+\frac{1}{g^{(00)}}\right)}, \quad \ell^{-2}_{0} = \frac{g^{(1-1)}}{2l^{2}\left(1+\frac{1}{g^{(00)}}\right)}, \nonumber\\
     &\alpha_{1}= \frac{\eta^{2}_{v}(1+2\eta_{H})}{2\left(\frac{1}{g^{(1-1)}}+\frac{1}{g^{(-11)}}+1\right)},  \ell^{-2}_{1} = \frac{g^{(00)}}{2l^{2}\left(\frac{1}{g^{(1-1)}}+\frac{1}{g^{(-11)}}+1\right)},\nonumber\\
      &\alpha_{2}= -\frac{\eta^{2}_{v}(1+2\eta_{H})}{2\left(1+\frac{1}{g^{(00)}}\right)}, \quad \ell^{-2}_{2} = \frac{g^{(-11)}}{2l^{2}\left(1+\frac{1}{g^{(00)}}\right)} 
\end{align}

In Fig.~\ref{fig:delsigmaB_elastic-mag-z} (a), we plot the corresponding magnetoconductivity as a function of the parameter $a$. When $a=0$, we observe WAL, as expected for graphene. Remarkably, as the parameter $a$ is increased, WAL is notably enhanced even though the Berry phase decreases from its peak value $\pi$ at $a=0$. As the parameter $a$ is increased to $2^{-\frac{1}{2}}$, WAL gradually crossovers to WL. The $\alpha\rightarrow \alpha^{-1}$ duality of the $\alpha-\mathcal{T}_3$ is reflected in the weak (anti)localization as $a$ is increased beyond $2^{-\frac{1}{2}}$. The notable enhancement in WAL is attributed to the presence of two WAL channels ($g^{(-11)}$ and $g^{(1-1)}$) when $0<a<1$. We estimate that WAL is maximum when $a=1/4$ or when $a=\sqrt{3}/2$.

In the presence of only magnetic $S_z$ impurities ($\alpha_z=1$), the Cooperon gaps $g^{(-22)}$ and $g^{(2-2)}$ yield the diverging contributions. In this case, we evaluate the magnetoconductivity to be~\cite{SI} 
\begin{align}
    &\Delta\sigma(B) =\nonumber\\
    &\frac{e^{2}}{\pi h}\sum_{i=0, 1}\alpha_{i}\left[\Psi\left(\frac{\ell^{2}_{B}}{\ell^{2}_{\phi}}+\frac{\ell^{2}_{B}}{\ell^{2}_{i}}+\frac{1}{2}\right)-\ln\left(\frac{\ell^{2}_{B}}{\ell^{2}_{\phi}}+\frac{\ell^{2}_{B}}{\ell^{2}_{i}}\right)\right],
\end{align}
where 
\begin{align}
      \alpha_{0}= \frac{1}{2}, \quad \alpha_{1}= \frac{1}{2},\quad
     \ell^{-2}_{0}=\frac{g^{(2-2)}}{2\ell^{2}} , \quad \ell^{-2}_{1} =\frac{g^{(-22)}}{2\ell^{2}}.
\end{align}

In Fig.~\ref{fig:delsigmaB_elastic-mag-z} (b), we plot the corresponding magnetoconductivity. We observe WL throughout that attains its maximum value for the dice lattice. For only $S_x$ impurities (i.e. $\alpha_x=1/2$), no diverging contributions to the Cooperon channels are obtained, and hence we do not discuss this case here. 

Quantum interference of electrons confined in the two-dimensional $\alpha-\mathcal{T}_3$ lattice can result in strikingly different localization-antilocalization properties, which can be manipulated by controlling the hopping strength of the electrons as well by magnetic doping. 
Our work makes an important advance in the study of electron transport in two-dimensional materials. 

\bibliography{biblio.bib}

\begin{thebibliography}{27}%
\makeatletter
\providecommand \@ifxundefined [1]{%
 \@ifx{#1\undefined}
}%
\providecommand \@ifnum [1]{%
 \ifnum #1\expandafter \@firstoftwo
 \else \expandafter \@secondoftwo
 \fi
}%
\providecommand \@ifx [1]{%
 \ifx #1\expandafter \@firstoftwo
 \else \expandafter \@secondoftwo
 \fi
}%
\providecommand \natexlab [1]{#1}%
\providecommand \enquote  [1]{``#1''}%
\providecommand \bibnamefont  [1]{#1}%
\providecommand \bibfnamefont [1]{#1}%
\providecommand \citenamefont [1]{#1}%
\providecommand \href@noop [0]{\@secondoftwo}%
\providecommand \href [0]{\begingroup \@sanitize@url \@href}%
\providecommand \@href[1]{\@@startlink{#1}\@@href}%
\providecommand \@@href[1]{\endgroup#1\@@endlink}%
\providecommand \@sanitize@url [0]{\catcode `\\12\catcode `\$12\catcode
  `\&12\catcode `\#12\catcode `\^12\catcode `\_12\catcode `\%12\relax}%
\providecommand \@@startlink[1]{}%
\providecommand \@@endlink[0]{}%
\providecommand \url  [0]{\begingroup\@sanitize@url \@url }%
\providecommand \@url [1]{\endgroup\@href {#1}{\urlprefix }}%
\providecommand \urlprefix  [0]{URL }%
\providecommand \Eprint [0]{\href }%
\providecommand \doibase [0]{https://doi.org/}%
\providecommand \selectlanguage [0]{\@gobble}%
\providecommand \bibinfo  [0]{\@secondoftwo}%
\providecommand \bibfield  [0]{\@secondoftwo}%
\providecommand \translation [1]{[#1]}%
\providecommand \BibitemOpen [0]{}%
\providecommand \bibitemStop [0]{}%
\providecommand \bibitemNoStop [0]{.\EOS\space}%
\providecommand \EOS [0]{\spacefactor3000\relax}%
\providecommand \BibitemShut  [1]{\csname bibitem#1\endcsname}%
\let\auto@bib@innerbib\@empty
\bibitem [{\citenamefont {Anderson}(1958)}]{anderson1958absence}%
  \BibitemOpen
  \bibfield  {author} {\bibinfo {author} {\bibfnamefont {P.~W.}\ \bibnamefont
  {Anderson}},\ }\bibfield  {title} {\bibinfo {title} {Absence of diffusion in
  certain random lattices},\ }\href@noop {} {\bibfield  {journal} {\bibinfo
  {journal} {Physical review}\ }\textbf {\bibinfo {volume} {109}},\ \bibinfo
  {pages} {1492} (\bibinfo {year} {1958})}\BibitemShut {NoStop}%
\bibitem [{\citenamefont {Altshuler}\ \emph {et~al.}(1980)\citenamefont
  {Altshuler}, \citenamefont {Khmel'Nitzkii}, \citenamefont {Larkin},\ and\
  \citenamefont {Lee}}]{altshuler1980magnetoresistance}%
  \BibitemOpen
  \bibfield  {author} {\bibinfo {author} {\bibfnamefont {B.}~\bibnamefont
  {Altshuler}}, \bibinfo {author} {\bibfnamefont {D.}~\bibnamefont
  {Khmel'Nitzkii}}, \bibinfo {author} {\bibfnamefont {A.}~\bibnamefont
  {Larkin}},\ and\ \bibinfo {author} {\bibfnamefont {P.}~\bibnamefont {Lee}},\
  }\bibfield  {title} {\bibinfo {title} {Magnetoresistance and hall effect in a
  disordered two-dimensional electron gas},\ }\href@noop {} {\bibfield
  {journal} {\bibinfo  {journal} {Physical Review B}\ }\textbf {\bibinfo
  {volume} {22}},\ \bibinfo {pages} {5142} (\bibinfo {year}
  {1980})}\BibitemShut {NoStop}%
\bibitem [{\citenamefont {Bergmann}(1984)}]{bergmann1984weak}%
  \BibitemOpen
  \bibfield  {author} {\bibinfo {author} {\bibfnamefont {G.}~\bibnamefont
  {Bergmann}},\ }\bibfield  {title} {\bibinfo {title} {Weak localization in
  thin films: a time-of-flight experiment with conduction electrons},\
  }\href@noop {} {\bibfield  {journal} {\bibinfo  {journal} {Physics Reports}\
  }\textbf {\bibinfo {volume} {107}},\ \bibinfo {pages} {1} (\bibinfo {year}
  {1984})}\BibitemShut {NoStop}%
\bibitem [{\citenamefont {Chakravarty}\ and\ \citenamefont
  {Schmid}(1986)}]{chakravarty1986weak}%
  \BibitemOpen
  \bibfield  {author} {\bibinfo {author} {\bibfnamefont {S.}~\bibnamefont
  {Chakravarty}}\ and\ \bibinfo {author} {\bibfnamefont {A.}~\bibnamefont
  {Schmid}},\ }\bibfield  {title} {\bibinfo {title} {Weak localization: The
  quasiclassical theory of electrons in a random potential},\ }\href@noop {}
  {\bibfield  {journal} {\bibinfo  {journal} {Physics Reports}\ }\textbf
  {\bibinfo {volume} {140}},\ \bibinfo {pages} {193} (\bibinfo {year}
  {1986})}\BibitemShut {NoStop}%
\bibitem [{\citenamefont {Beenakker}\ and\ \citenamefont {van
  Houten}(1991)}]{beenakker1991quantum}%
  \BibitemOpen
  \bibfield  {author} {\bibinfo {author} {\bibfnamefont {C.}~\bibnamefont
  {Beenakker}}\ and\ \bibinfo {author} {\bibfnamefont {H.}~\bibnamefont {van
  Houten}},\ }\bibfield  {title} {\bibinfo {title} {Quantum transport in
  semiconductor nanostructures},\ }in\ \href@noop {} {\emph {\bibinfo
  {booktitle} {Solid state physics}}},\ Vol.~\bibinfo {volume} {44}\ (\bibinfo
  {publisher} {Elsevier},\ \bibinfo {year} {1991})\ pp.\ \bibinfo {pages}
  {1--228}\BibitemShut {NoStop}%
\bibitem [{\citenamefont {Hikami}\ \emph {et~al.}(1980)\citenamefont {Hikami},
  \citenamefont {Larkin},\ and\ \citenamefont {Nagaoka}}]{hikami1980spin}%
  \BibitemOpen
  \bibfield  {author} {\bibinfo {author} {\bibfnamefont {S.}~\bibnamefont
  {Hikami}}, \bibinfo {author} {\bibfnamefont {A.~I.}\ \bibnamefont {Larkin}},\
  and\ \bibinfo {author} {\bibfnamefont {Y.}~\bibnamefont {Nagaoka}},\
  }\bibfield  {title} {\bibinfo {title} {Spin-orbit interaction and
  magnetoresistance in the two dimensional random system},\ }\href@noop {}
  {\bibfield  {journal} {\bibinfo  {journal} {Progress of Theoretical Physics}\
  }\textbf {\bibinfo {volume} {63}},\ \bibinfo {pages} {707} (\bibinfo {year}
  {1980})}\BibitemShut {NoStop}%
\bibitem [{\citenamefont {Suzuura}\ and\ \citenamefont
  {Ando}(2002)}]{suzuura2002crossover}%
  \BibitemOpen
  \bibfield  {author} {\bibinfo {author} {\bibfnamefont {H.}~\bibnamefont
  {Suzuura}}\ and\ \bibinfo {author} {\bibfnamefont {T.}~\bibnamefont {Ando}},\
  }\bibfield  {title} {\bibinfo {title} {Crossover from symplectic to
  orthogonal class in a two-dimensional honeycomb lattice},\ }\href@noop {}
  {\bibfield  {journal} {\bibinfo  {journal} {Physical Review Letters}\
  }\textbf {\bibinfo {volume} {89}},\ \bibinfo {pages} {266603} (\bibinfo
  {year} {2002})}\BibitemShut {NoStop}%
\bibitem [{\citenamefont {Khveshchenko}(2006)}]{khveshchenko2006electron}%
  \BibitemOpen
  \bibfield  {author} {\bibinfo {author} {\bibfnamefont {D.}~\bibnamefont
  {Khveshchenko}},\ }\bibfield  {title} {\bibinfo {title} {Electron
  localization properties in graphene},\ }\href@noop {} {\bibfield  {journal}
  {\bibinfo  {journal} {Physical Review Letters}\ }\textbf {\bibinfo {volume}
  {97}},\ \bibinfo {pages} {036802} (\bibinfo {year} {2006})}\BibitemShut
  {NoStop}%
\bibitem [{\citenamefont {McCann}\ \emph {et~al.}(2006)\citenamefont {McCann},
  \citenamefont {Kechedzhi}, \citenamefont {Fal’ko}, \citenamefont {Suzuura},
  \citenamefont {Ando},\ and\ \citenamefont {Altshuler}}]{mccann2006weak}%
  \BibitemOpen
  \bibfield  {author} {\bibinfo {author} {\bibfnamefont {E.}~\bibnamefont
  {McCann}}, \bibinfo {author} {\bibfnamefont {K.}~\bibnamefont {Kechedzhi}},
  \bibinfo {author} {\bibfnamefont {V.~I.}\ \bibnamefont {Fal’ko}}, \bibinfo
  {author} {\bibfnamefont {H.}~\bibnamefont {Suzuura}}, \bibinfo {author}
  {\bibfnamefont {T.}~\bibnamefont {Ando}},\ and\ \bibinfo {author}
  {\bibfnamefont {B.}~\bibnamefont {Altshuler}},\ }\bibfield  {title} {\bibinfo
  {title} {Weak-localization magnetoresistance and valley symmetry in
  graphene},\ }\href@noop {} {\bibfield  {journal} {\bibinfo  {journal}
  {Physical Review Letters}\ }\textbf {\bibinfo {volume} {97}},\ \bibinfo
  {pages} {146805} (\bibinfo {year} {2006})}\BibitemShut {NoStop}%
\bibitem [{\citenamefont {Gorbachev}\ \emph {et~al.}(2007)\citenamefont
  {Gorbachev}, \citenamefont {Tikhonenko}, \citenamefont {Mayorov},
  \citenamefont {Horsell},\ and\ \citenamefont
  {Savchenko}}]{gorbachev2007weak}%
  \BibitemOpen
  \bibfield  {author} {\bibinfo {author} {\bibfnamefont {R.}~\bibnamefont
  {Gorbachev}}, \bibinfo {author} {\bibfnamefont {F.}~\bibnamefont
  {Tikhonenko}}, \bibinfo {author} {\bibfnamefont {A.}~\bibnamefont {Mayorov}},
  \bibinfo {author} {\bibfnamefont {D.}~\bibnamefont {Horsell}},\ and\ \bibinfo
  {author} {\bibfnamefont {A.}~\bibnamefont {Savchenko}},\ }\bibfield  {title}
  {\bibinfo {title} {Weak localization in bilayer graphene},\ }\href@noop {}
  {\bibfield  {journal} {\bibinfo  {journal} {Physical Review Letters}\
  }\textbf {\bibinfo {volume} {98}},\ \bibinfo {pages} {176805} (\bibinfo
  {year} {2007})}\BibitemShut {NoStop}%
\bibitem [{\citenamefont {Wu}\ \emph {et~al.}(2007)\citenamefont {Wu},
  \citenamefont {Li}, \citenamefont {Song}, \citenamefont {Berger},\ and\
  \citenamefont {de~Heer}}]{wu2007weak}%
  \BibitemOpen
  \bibfield  {author} {\bibinfo {author} {\bibfnamefont {X.}~\bibnamefont
  {Wu}}, \bibinfo {author} {\bibfnamefont {X.}~\bibnamefont {Li}}, \bibinfo
  {author} {\bibfnamefont {Z.}~\bibnamefont {Song}}, \bibinfo {author}
  {\bibfnamefont {C.}~\bibnamefont {Berger}},\ and\ \bibinfo {author}
  {\bibfnamefont {W.~A.}\ \bibnamefont {de~Heer}},\ }\bibfield  {title}
  {\bibinfo {title} {Weak antilocalization in epitaxial graphene: Evidence for
  chiral electrons},\ }\href@noop {} {\bibfield  {journal} {\bibinfo  {journal}
  {Physical Review Letters}\ }\textbf {\bibinfo {volume} {98}},\ \bibinfo
  {pages} {136801} (\bibinfo {year} {2007})}\BibitemShut {NoStop}%
\bibitem [{\citenamefont {Tikhonenko}\ \emph {et~al.}(2008)\citenamefont
  {Tikhonenko}, \citenamefont {Horsell}, \citenamefont {Gorbachev},\ and\
  \citenamefont {Savchenko}}]{tikhonenko2008weak}%
  \BibitemOpen
  \bibfield  {author} {\bibinfo {author} {\bibfnamefont {F.}~\bibnamefont
  {Tikhonenko}}, \bibinfo {author} {\bibfnamefont {D.}~\bibnamefont {Horsell}},
  \bibinfo {author} {\bibfnamefont {R.}~\bibnamefont {Gorbachev}},\ and\
  \bibinfo {author} {\bibfnamefont {A.}~\bibnamefont {Savchenko}},\ }\bibfield
  {title} {\bibinfo {title} {Weak localization in graphene flakes},\
  }\href@noop {} {\bibfield  {journal} {\bibinfo  {journal} {Physical Review
  Letters}\ }\textbf {\bibinfo {volume} {100}},\ \bibinfo {pages} {056802}
  (\bibinfo {year} {2008})}\BibitemShut {NoStop}%
\bibitem [{\citenamefont {Tkachov}\ and\ \citenamefont
  {Hankiewicz}(2011)}]{tkachov2011weak}%
  \BibitemOpen
  \bibfield  {author} {\bibinfo {author} {\bibfnamefont {G.}~\bibnamefont
  {Tkachov}}\ and\ \bibinfo {author} {\bibfnamefont {E.}~\bibnamefont
  {Hankiewicz}},\ }\bibfield  {title} {\bibinfo {title} {Weak antilocalization
  in hgte quantum wells and topological surface states: Massive versus massless
  dirac fermions},\ }\href@noop {} {\bibfield  {journal} {\bibinfo  {journal}
  {Physical Review B}\ }\textbf {\bibinfo {volume} {84}},\ \bibinfo {pages}
  {035444} (\bibinfo {year} {2011})}\BibitemShut {NoStop}%
\bibitem [{\citenamefont {Lu}\ \emph {et~al.}(2011)\citenamefont {Lu},
  \citenamefont {Shi},\ and\ \citenamefont {Shen}}]{lu2011competition}%
  \BibitemOpen
  \bibfield  {author} {\bibinfo {author} {\bibfnamefont {H.-Z.}\ \bibnamefont
  {Lu}}, \bibinfo {author} {\bibfnamefont {J.}~\bibnamefont {Shi}},\ and\
  \bibinfo {author} {\bibfnamefont {S.-Q.}\ \bibnamefont {Shen}},\ }\bibfield
  {title} {\bibinfo {title} {Competition between weak localization and
  antilocalization in topological surface states},\ }\href@noop {} {\bibfield
  {journal} {\bibinfo  {journal} {Physical Review Letters}\ }\textbf {\bibinfo
  {volume} {107}},\ \bibinfo {pages} {076801} (\bibinfo {year}
  {2011})}\BibitemShut {NoStop}%
\bibitem [{\citenamefont {Lu}\ \emph {et~al.}(2013)\citenamefont {Lu},
  \citenamefont {Yao}, \citenamefont {Xiao},\ and\ \citenamefont
  {Shen}}]{lu2013intervalley}%
  \BibitemOpen
  \bibfield  {author} {\bibinfo {author} {\bibfnamefont {H.-Z.}\ \bibnamefont
  {Lu}}, \bibinfo {author} {\bibfnamefont {W.}~\bibnamefont {Yao}}, \bibinfo
  {author} {\bibfnamefont {D.}~\bibnamefont {Xiao}},\ and\ \bibinfo {author}
  {\bibfnamefont {S.-Q.}\ \bibnamefont {Shen}},\ }\bibfield  {title} {\bibinfo
  {title} {Intervalley scattering and localization behaviors of spin-valley
  coupled dirac fermions},\ }\href@noop {} {\bibfield  {journal} {\bibinfo
  {journal} {Physical Review Letters}\ }\textbf {\bibinfo {volume} {110}},\
  \bibinfo {pages} {016806} (\bibinfo {year} {2013})}\BibitemShut {NoStop}%
\bibitem [{\citenamefont {Lu}\ and\ \citenamefont {Shen}(2014)}]{lu2014finite}%
  \BibitemOpen
  \bibfield  {author} {\bibinfo {author} {\bibfnamefont {H.-Z.}\ \bibnamefont
  {Lu}}\ and\ \bibinfo {author} {\bibfnamefont {S.-Q.}\ \bibnamefont {Shen}},\
  }\bibfield  {title} {\bibinfo {title} {Finite-temperature conductivity and
  magnetoconductivity of topological insulators},\ }\href@noop {} {\bibfield
  {journal} {\bibinfo  {journal} {Physical Review Letters}\ }\textbf {\bibinfo
  {volume} {112}},\ \bibinfo {pages} {146601} (\bibinfo {year}
  {2014})}\BibitemShut {NoStop}%
\bibitem [{\citenamefont {Fu}\ \emph {et~al.}(2019)\citenamefont {Fu},
  \citenamefont {Wang},\ and\ \citenamefont {Shen}}]{fu2019quantum}%
  \BibitemOpen
  \bibfield  {author} {\bibinfo {author} {\bibfnamefont {B.}~\bibnamefont
  {Fu}}, \bibinfo {author} {\bibfnamefont {H.-W.}\ \bibnamefont {Wang}},\ and\
  \bibinfo {author} {\bibfnamefont {S.-Q.}\ \bibnamefont {Shen}},\ }\bibfield
  {title} {\bibinfo {title} {Quantum interference theory of magnetoresistance
  in dirac materials},\ }\href@noop {} {\bibfield  {journal} {\bibinfo
  {journal} {Physical Review Letters}\ }\textbf {\bibinfo {volume} {122}},\
  \bibinfo {pages} {246601} (\bibinfo {year} {2019})}\BibitemShut {NoStop}%
\bibitem [{\citenamefont {Neto}\ \emph {et~al.}(2009)\citenamefont {Neto},
  \citenamefont {Guinea}, \citenamefont {Peres}, \citenamefont {Novoselov},\
  and\ \citenamefont {Geim}}]{neto2009electronic}%
  \BibitemOpen
  \bibfield  {author} {\bibinfo {author} {\bibfnamefont {A.~C.}\ \bibnamefont
  {Neto}}, \bibinfo {author} {\bibfnamefont {F.}~\bibnamefont {Guinea}},
  \bibinfo {author} {\bibfnamefont {N.~M.}\ \bibnamefont {Peres}}, \bibinfo
  {author} {\bibfnamefont {K.~S.}\ \bibnamefont {Novoselov}},\ and\ \bibinfo
  {author} {\bibfnamefont {A.~K.}\ \bibnamefont {Geim}},\ }\bibfield  {title}
  {\bibinfo {title} {The electronic properties of graphene},\ }\href@noop {}
  {\bibfield  {journal} {\bibinfo  {journal} {Reviews of modern physics}\
  }\textbf {\bibinfo {volume} {81}},\ \bibinfo {pages} {109} (\bibinfo {year}
  {2009})}\BibitemShut {NoStop}%
\bibitem [{\citenamefont {Sarma}\ \emph {et~al.}(2011)\citenamefont {Sarma},
  \citenamefont {Adam}, \citenamefont {Hwang},\ and\ \citenamefont
  {Rossi}}]{sarma2011electronic}%
  \BibitemOpen
  \bibfield  {author} {\bibinfo {author} {\bibfnamefont {S.~D.}\ \bibnamefont
  {Sarma}}, \bibinfo {author} {\bibfnamefont {S.}~\bibnamefont {Adam}},
  \bibinfo {author} {\bibfnamefont {E.}~\bibnamefont {Hwang}},\ and\ \bibinfo
  {author} {\bibfnamefont {E.}~\bibnamefont {Rossi}},\ }\bibfield  {title}
  {\bibinfo {title} {Electronic transport in two-dimensional graphene},\
  }\href@noop {} {\bibfield  {journal} {\bibinfo  {journal} {Reviews of modern
  physics}\ }\textbf {\bibinfo {volume} {83}},\ \bibinfo {pages} {407}
  (\bibinfo {year} {2011})}\BibitemShut {NoStop}%
\bibitem [{\citenamefont {Armitage}\ \emph {et~al.}(2018)\citenamefont
  {Armitage}, \citenamefont {Mele},\ and\ \citenamefont
  {Vishwanath}}]{armitage2018weyl}%
  \BibitemOpen
  \bibfield  {author} {\bibinfo {author} {\bibfnamefont {N.}~\bibnamefont
  {Armitage}}, \bibinfo {author} {\bibfnamefont {E.}~\bibnamefont {Mele}},\
  and\ \bibinfo {author} {\bibfnamefont {A.}~\bibnamefont {Vishwanath}},\
  }\bibfield  {title} {\bibinfo {title} {Weyl and dirac semimetals in
  three-dimensional solids},\ }\href@noop {} {\bibfield  {journal} {\bibinfo
  {journal} {Reviews of Modern Physics}\ }\textbf {\bibinfo {volume} {90}},\
  \bibinfo {pages} {015001} (\bibinfo {year} {2018})}\BibitemShut {NoStop}%
\bibitem [{\citenamefont {Cao}\ \emph {et~al.}(2018)\citenamefont {Cao},
  \citenamefont {Fatemi}, \citenamefont {Fang}, \citenamefont {Watanabe},
  \citenamefont {Taniguchi}, \citenamefont {Kaxiras},\ and\ \citenamefont
  {Jarillo-Herrero}}]{cao2018unconventional}%
  \BibitemOpen
  \bibfield  {author} {\bibinfo {author} {\bibfnamefont {Y.}~\bibnamefont
  {Cao}}, \bibinfo {author} {\bibfnamefont {V.}~\bibnamefont {Fatemi}},
  \bibinfo {author} {\bibfnamefont {S.}~\bibnamefont {Fang}}, \bibinfo {author}
  {\bibfnamefont {K.}~\bibnamefont {Watanabe}}, \bibinfo {author}
  {\bibfnamefont {T.}~\bibnamefont {Taniguchi}}, \bibinfo {author}
  {\bibfnamefont {E.}~\bibnamefont {Kaxiras}},\ and\ \bibinfo {author}
  {\bibfnamefont {P.}~\bibnamefont {Jarillo-Herrero}},\ }\bibfield  {title}
  {\bibinfo {title} {Unconventional superconductivity in magic-angle graphene
  superlattices},\ }\href@noop {} {\bibfield  {journal} {\bibinfo  {journal}
  {Nature}\ }\textbf {\bibinfo {volume} {556}},\ \bibinfo {pages} {43}
  (\bibinfo {year} {2018})}\BibitemShut {NoStop}%
\bibitem [{\citenamefont {Raoux}\ \emph {et~al.}(2014)\citenamefont {Raoux},
  \citenamefont {Morigi}, \citenamefont {Fuchs}, \citenamefont {Pi{\'e}chon},\
  and\ \citenamefont {Montambaux}}]{raoux2014dia}%
  \BibitemOpen
  \bibfield  {author} {\bibinfo {author} {\bibfnamefont {A.}~\bibnamefont
  {Raoux}}, \bibinfo {author} {\bibfnamefont {M.}~\bibnamefont {Morigi}},
  \bibinfo {author} {\bibfnamefont {J.-N.}\ \bibnamefont {Fuchs}}, \bibinfo
  {author} {\bibfnamefont {F.}~\bibnamefont {Pi{\'e}chon}},\ and\ \bibinfo
  {author} {\bibfnamefont {G.}~\bibnamefont {Montambaux}},\ }\bibfield  {title}
  {\bibinfo {title} {From dia-to paramagnetic orbital susceptibility of
  massless fermions},\ }\href@noop {} {\bibfield  {journal} {\bibinfo
  {journal} {Physical Review Letters}\ }\textbf {\bibinfo {volume} {112}},\
  \bibinfo {pages} {026402} (\bibinfo {year} {2014})}\BibitemShut {NoStop}%
\bibitem [{\citenamefont {Wang}\ and\ \citenamefont
  {Ran}(2011)}]{wang2011nearly}%
  \BibitemOpen
  \bibfield  {author} {\bibinfo {author} {\bibfnamefont {F.}~\bibnamefont
  {Wang}}\ and\ \bibinfo {author} {\bibfnamefont {Y.}~\bibnamefont {Ran}},\
  }\bibfield  {title} {\bibinfo {title} {Nearly flat band with chern number c=
  2 on the dice lattice},\ }\href@noop {} {\bibfield  {journal} {\bibinfo
  {journal} {Physical Review B}\ }\textbf {\bibinfo {volume} {84}},\ \bibinfo
  {pages} {241103} (\bibinfo {year} {2011})}\BibitemShut {NoStop}%
\bibitem [{\citenamefont {Rizzi}\ \emph {et~al.}(2006)\citenamefont {Rizzi},
  \citenamefont {Cataudella},\ and\ \citenamefont {Fazio}}]{rizzi2006phase}%
  \BibitemOpen
  \bibfield  {author} {\bibinfo {author} {\bibfnamefont {M.}~\bibnamefont
  {Rizzi}}, \bibinfo {author} {\bibfnamefont {V.}~\bibnamefont {Cataudella}},\
  and\ \bibinfo {author} {\bibfnamefont {R.}~\bibnamefont {Fazio}},\ }\bibfield
   {title} {\bibinfo {title} {Phase diagram of the bose-hubbard model with t 3
  symmetry},\ }\href@noop {} {\bibfield  {journal} {\bibinfo  {journal}
  {Physical Review B}\ }\textbf {\bibinfo {volume} {73}},\ \bibinfo {pages}
  {144511} (\bibinfo {year} {2006})}\BibitemShut {NoStop}%
\bibitem [{\citenamefont {Malcolm}\ and\ \citenamefont
  {Nicol}(2015)}]{malcolm2015magneto}%
  \BibitemOpen
  \bibfield  {author} {\bibinfo {author} {\bibfnamefont {J.~D.}\ \bibnamefont
  {Malcolm}}\ and\ \bibinfo {author} {\bibfnamefont {E.~J.}\ \bibnamefont
  {Nicol}},\ }\bibfield  {title} {\bibinfo {title} {Magneto-optics of massless
  kane fermions: Role of the flat band and unusual berry phase},\ }\href@noop
  {} {\bibfield  {journal} {\bibinfo  {journal} {Physical Review B}\ }\textbf
  {\bibinfo {volume} {92}},\ \bibinfo {pages} {035118} (\bibinfo {year}
  {2015})}\BibitemShut {NoStop}%
\bibitem [{\citenamefont {Serret}\ \emph {et~al.}(2002)\citenamefont {Serret},
  \citenamefont {Butaud},\ and\ \citenamefont {Pannetier}}]{serret2002vortex}%
  \BibitemOpen
  \bibfield  {author} {\bibinfo {author} {\bibfnamefont {E.}~\bibnamefont
  {Serret}}, \bibinfo {author} {\bibfnamefont {P.}~\bibnamefont {Butaud}},\
  and\ \bibinfo {author} {\bibfnamefont {B.}~\bibnamefont {Pannetier}},\
  }\bibfield  {title} {\bibinfo {title} {Vortex correlations in a fully
  frustrated two-dimensional superconducting network},\ }\href@noop {}
  {\bibfield  {journal} {\bibinfo  {journal} {EPL (Europhysics Letters)}\
  }\textbf {\bibinfo {volume} {59}},\ \bibinfo {pages} {225} (\bibinfo {year}
  {2002})}\BibitemShut {NoStop}%
\bibitem [{SI(2023)}]{SI}%
  \BibitemOpen
  \bibfield  {title} {\bibinfo {title} {See supplementary information},\
  }\href@noop {} {\  (\bibinfo {year} {2023})}\BibitemShut {NoStop}%
\end{thebibliography}%
\clearpage
\begin{widetext}
\huge{\centering\section*{
Supplemental Information}}
\normalsize
\section{Model}
Pseudospin-1 fermions in the  $\alpha-\mathcal{T}_3$ lattice model can be described by the momentum space Hamiltonian
\begin{equation}
 H(\mathbf{k}|\mu)=\hbar v_{F}
 \begin{pmatrix}
 0& a f_{\mu}(\mathbf{k}) & 0\\
 a f_{\mu}^{*}(\mathbf{k}) & 0&  b f_{\mu}(\mathbf{k})\\
 0 &  b f^{*}_{\mu}{ (\mathbf{k})} &0
\end{pmatrix}
\label{Eq_1}
\end{equation}
\begin{figure}[h]
    \centering
    \includegraphics[width=0.55\textwidth]{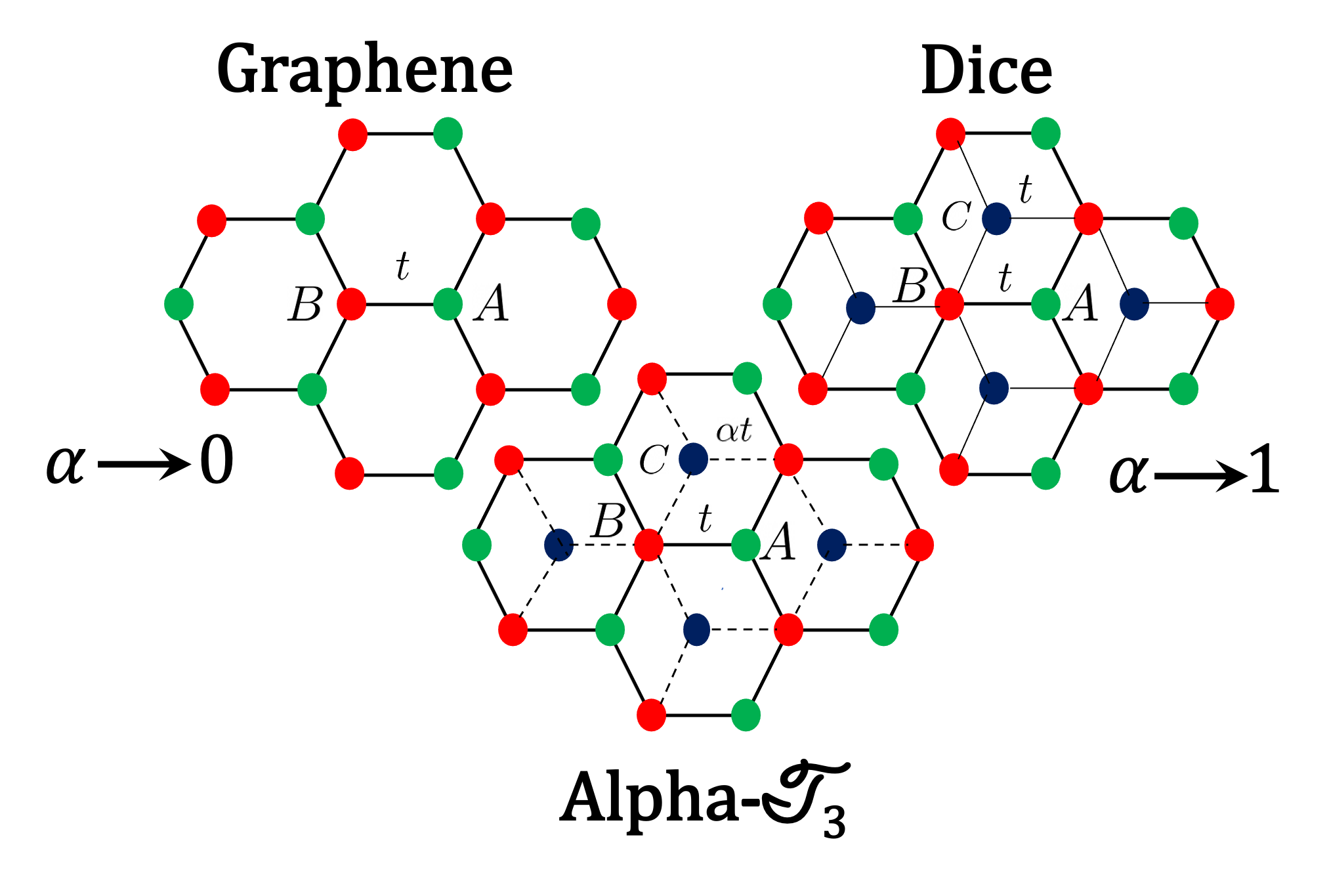}
    \caption{The lattice of the $\alpha-T_3$ model with hopping $t$ between the sublattices $A$ and $B$ and  $\alpha t$ between the sublattices at the $B$ and $C$.}
    \label{fig:1}
\end{figure}
where $f_{\mu}(\mathbf{k})=\mu k_{x}-ik_{y}$ with $\mu=\pm1$ is the valley index for positive and negative valleys, respectively. The wave vector is $(k_{x},k_{y})$, $\hbar$ is the reduced Planck's constant, $v_{F}$ is the Fermi velocity. $\psi=\tan^{1}(\alpha)$ and $a=\cos\psi$ and $b=\sin \psi$ with $a^{2}+b^{2}=1$.

\section{Diagrammatic techniques}
Here we calculate various  quantities using the Feynman diagram technique to evaluate the quantum interference correction to the conductivity.
\subsection{Eigenstates}
The Hamiltonian in Eq.~\ref{Eq_1} describes pseudospin-1 fermions in two dimensions with a conduction and a valence band for both valleys. We assume that the Fermi level intersects with the conduction band, and in the weak scattering limit $E_{F}\ll{\hbar}/{\tau}$, the valence band becomes irrelevant for transport, where $\tau$ is the total scattering time. The energy dispersion for the conduction band is 
\begin{equation}
\epsilon_{\textbf{k}}=\hbar v_{F}\sqrt{k^{2}_{x}+k^{2}_{y}}=\hbar v_{F}k
\label{Eq_2}
\end{equation}
\begin{figure}[h]
    \centering
    \includegraphics[width=0.35\textwidth]{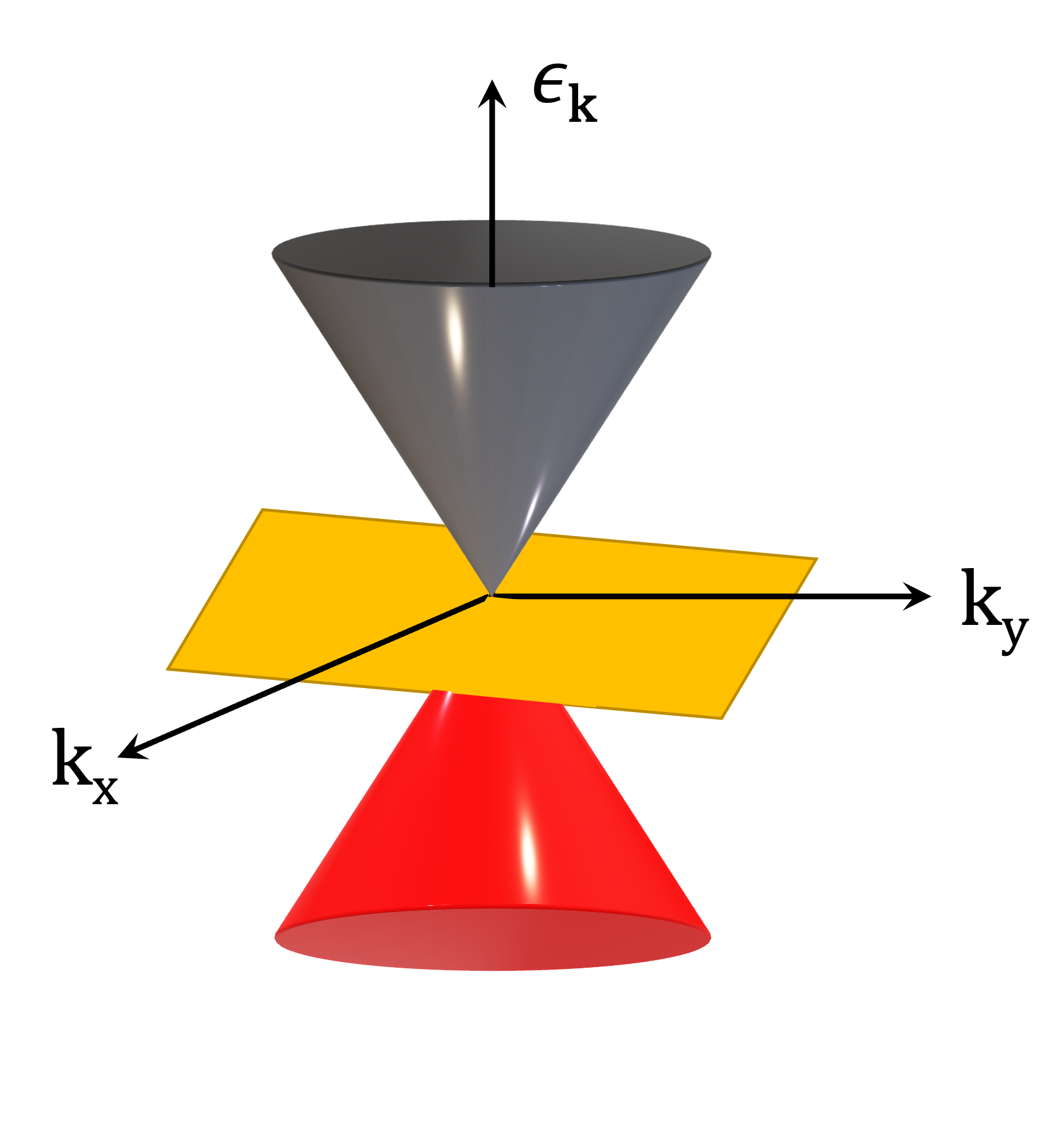}
    \caption{A schematic diagram of the energy dispersion near a $\mathbf{K}$ point, with linearly dispersing valence and conduction bands, and a dispersionless flat band at zero energy that cuts through the Dirac point.}
    \label{fig:2}
\end{figure}
In two dimensions $k_{x}=k\cos\phi$, $k_{y}=k\sin\phi$, where $\tan\phi={k_{y}}/{k_{x}}$; its state is described by
\begin{equation}
\ket{\textbf{k}, \mu}=\frac{1}{\sqrt{2}}
\begin{pmatrix}
   \mu ae^{-i\mu\phi}\\
   1\\
   \mu be^{i\mu\phi}
   \end{pmatrix}e^{i\textbf{k}\cdot\textbf{r}} \\
 \end{equation}
 for the positive valley, we have
 \begin{equation}
 \ket{\textbf{k}, +}=\frac{1}{\sqrt{2}}
\begin{pmatrix}
    ae^{-i\phi}\\
   1\\
    be^{i\phi}
   \end{pmatrix}e^{i\textbf{k}\cdot\textbf{r}} \\
 \end{equation}    
$a=\cos\psi$, $b=\sin\psi$ and $\tan\phi=\frac{k_{y}}{k_{x}}$. The Fermi energy $E_{F}$ is measured from the node.
\subsection{Impurity potentials}
The impurity potentials are given by,
\begin{equation}
    U(\textbf{r})=U_{0}(\textbf{r})+U_{m}(\textbf{r})
\end{equation}
where $U_{0}(\textbf{r})$ is for the elastic scattering, $U_{m}(\textbf{r})$ is for the magnetic scattering and
\begin{align}
 &U_{0}(\textbf{r})=\sum_{i}u^{i}_{0}\delta(\textbf{r}-\textbf{R}_{i}), \nonumber\\
 &U_{m}(\textbf{r})=\sum_{i}\sum_{\alpha=x,y,z}u^{i}_{\alpha}S_{\alpha}\delta(\textbf{r}-\textbf{R}_{i}),
\end{align}
where $u(\textbf{r}-\textbf{R}_{i})$ represents the random potential by an impurity located at $\textbf{R}_{i}$. $\Vec{S}=(S_{x},S_{y},S_{z})$ is the vector of spin-1 matrices and 
\begin{align}
   S_{x} = \begin{pmatrix}
0 & 1 & 0\\
1 & 0 & 1\\
0 & 1 & 0
\end{pmatrix}, S_{y} = \begin{pmatrix}
0 & -i & 0\\
i & 0 & -i\\
0 & i & 0
\end{pmatrix}, S_{z} = \begin{pmatrix}
1 & 0 & 0\\
0 & 0 & 0\\
0 & 0 & -1
\end{pmatrix}.
\end{align}
The scattering (Born) amplitude $U_{\textbf{k},\textbf{k}'}$ can be found as
\begin{equation}
      U_{\textbf{k},\textbf{k}'} \equiv \bra{\textbf{k},+} U(\textbf{r})\ket{\textbf{k}',+}
= \bra{\textbf{k},+} U_{0}(\textbf{r})\ket{\textbf{k}',+}+\bra{\textbf{k},+} U_{m}(\textbf{r})\ket{\textbf{k}',+}
\end{equation}
For the elastic scattering

    \begin{align}
        &\bra{\textbf{k},+} U_{0}(\textbf{r})\ket{\textbf{k}',+} = U^{0}_{\mathbf{k,k'}}\nonumber\\
        &= \frac{1}{\sqrt{2}}\begin{pmatrix}
            ae^{i\phi} & 1 &  be^{-i\phi}
        \end{pmatrix}\sum_{i}u^{i}_{0}\delta(\mathbf{r-R_{i}})
       \frac{1}{\sqrt{2}} \begin{pmatrix}
              ae^{-i\phi'}  \\
              1 \\
               be^{i\phi'}
            \end{pmatrix}e^{(\mathbf{k'-k}).\mathbf{r}} \nonumber\\
        &=\frac{1}{2}\sum_{i}u^{i}_{0}e^{(\textbf{k}'-\textbf{k}).\textbf{R}_{i}}\left[a^{2}e^{i(\phi-\phi')}+ b^{2}e^{-i(\phi-\phi')}+1\right]
    \end{align}
    \begin{itemize}
        \item {For magnetic scattering}
    \end{itemize}
    \begin{equation}
          U_{m}(\mathbf{r})=\sum_{i}u^{i}_{x}S_{x}\delta(\textbf{r}-\textbf{R}_{i})+\sum_{i}u^{i}_{y}S_{y}\delta(\textbf{r}-\textbf{R}_{i})+\sum_{i}u^{i}_{z}S_{z}\delta(\textbf{r}-\textbf{R}_{i})
    \end{equation}
    where the Born amplitude for the $ x, y, z$ components of the magnetic impurity are
    \begin{align}
    &\bra{\textbf{k},
    +}\sum_{i}u^{i}_{x}S_{x}\delta(\textbf{r}-\textbf{R}_{i})\ket{\textbf{k}'+} = \nonumber\\  
    &\frac{1}{\sqrt{2}}\begin{pmatrix}
            ae^{i\phi} & 1 &  be^{-i\phi}
        \end{pmatrix}\sum_{i}u^{i}_{x}\delta(\textbf{r}-\textbf{R}_{i})\begin{pmatrix}
0 & 1 & 0\\
1 & 0 & 1\\
0 & 1 & 0
\end{pmatrix}
\frac{1}{\sqrt{2}}\begin{pmatrix}
              ae^{-i\phi'}  \\
              1 \\
               be^{i\phi'}
            \end{pmatrix}e^{(\mathbf{k'-k})\cdot\mathbf{r}} \nonumber\\
            &=
            \frac{1}{2}\sum_{i}u^{i}_{x}e^{(\mathbf{k'-k})\cdot\mathbf{R}_{i}}\begin{pmatrix}
            ae^{i\phi} & 1 &  be^{-i\phi}
        \end{pmatrix}\begin{pmatrix}
0 & 1 & 0\\
1 & 0 & 1\\
0 & 1 & 0
\end{pmatrix}\begin{pmatrix}
              ae^{-i\phi'}  \\
              1 \\
               be^{i\phi'}
            \end{pmatrix}
            \nonumber\\
            &= \frac{1}{2}\sum_{i}u^{i}_{x}e^{(\mathbf{k'-k}).\mathbf{R}_{i}}\left[a(e^{i\phi}+e^{-i\phi'})+b(e^{i\phi'}+e^{-i\phi})\right]
    \end{align}
\begin{align}
  &\bra{\textbf{k},
    +}\sum_{i}u^{i}_{y}S_{y}\delta(\textbf{r}-\textbf{R}_{i})\ket{\textbf{k}'+} \nonumber\\ &= \frac{1}{\sqrt{2}}\begin{pmatrix}
            ae^{i\phi} & 1 &  be^{-i\phi}
        \end{pmatrix}\sum_{i}u^{i}_{y}\delta(\textbf{r}-\textbf{R}_{i}) \begin{pmatrix}
0 & -i & 0\\
i & 0 & -i\\
0 & i & 0
\end{pmatrix}\frac{1}{\sqrt{2}} \begin{pmatrix}
              ae^{-i\phi'}  \\
              1 \\
               be^{i\phi'}
            \end{pmatrix}e^{(\mathbf{k'-k})\cdot\mathbf{r}} \nonumber\\ &
            = \frac{1}{2}\sum_{i}u^{i}_{y}e^{(\mathbf{k'-k})\cdot\mathbf{R}_{i}}\left[a(e^{-i\phi'}-e^{i\phi})+b(e^{-i\phi}-e^{i\phi'})\right]i
\end{align}
\begin{align}
     &\bra{\textbf{k},
    +}\sum_{i}u^{i}_{z}S_{z}\delta(\textbf{r}-\textbf{R}_{i})\ket{\textbf{k}'+} \nonumber \\ &= \frac{1}{\sqrt{2}}\begin{pmatrix}
            ae^{i\phi} & 1 &  be^{-i\phi}
        \end{pmatrix}\sum_{i}u^{i}_{z}\delta(\textbf{r}-\textbf{R}_{i}) \begin{pmatrix}
1 & 0 & 0\\
0 & 0 & 0\\
0 & 0 & -1
\end{pmatrix}\frac{1}{\sqrt{2}} \begin{pmatrix}
              ae^{-i\phi'}  \\
              1 \\
               be^{i\phi'}
            \end{pmatrix}e^{(\mathbf{k'-k})\cdot\mathbf{r}}
            \nonumber\\
            &= \frac{1}{2}\sum_{i}u^{i}_{z}e^{(\mathbf{k'-k}).\mathbf{R}_{i}}\left[a^{2}e^{i(\phi-\phi')}-b^{2}e^{-i(\phi-\phi')}\right].
\end{align}
Finally, the Born amplitude for the magnetic impurities is given as
\begin{align}
     &\bra{\textbf{k},+} U_{m}(\textbf{r})\ket{\textbf{k}',+} = \nonumber\\ &\frac{1}{2}\sum_{i}u^{i}_{0}e^{(\textbf{k}'-\textbf{k})\cdot\textbf{R}_{i}}
    \{u^{i}_{x}[a(e^{i\phi}+e^{-i\phi'})+b(e^{i\phi'}+e^{-i\phi})]
    + iu^{i}_{y}[a(e^{-i\phi'}-e^{i\phi})+b(e^{-i\phi}-e^{i\phi'})] \nonumber\\ &
    +u^{i}_{z}[a^{2}e^{i(\phi-\phi')}- b^{2}e^{-i(\phi-\phi')}]\}.
\end{align}
\subsection{Impurity correlation function }
The correlation between different types of scattering or different components of the same type is neglected, and the relaxation times can be given according to different scattering mechanisms (Mattheiessen's rule).
\begin{equation}
    \bigr \langle U_{\textbf{k},\textbf{k}'}U_{\textbf{k}',\textbf{k}}\bigl \rangle_{imp}=\bigr \langle U_{\textbf{k},\textbf{k}'}U_{\textbf{k}',\textbf{k}}\bigl \rangle_{e}+\bigr \langle U_{\textbf{k},\textbf{k}'}U_{\textbf{k}',\textbf{k}}\bigl \rangle_{m}
\end{equation}
where we have
\begin{equation}
    \bigr \langle U_{\textbf{k},\textbf{k}'}U_{\textbf{k}',\textbf{k}}\bigl \rangle_{m}=\bigr \langle U_{\textbf{k},\textbf{k}'}U_{\textbf{k}',\textbf{k}}\bigl \rangle_{x}+\bigr \langle U_{\textbf{k},\textbf{k}'}U_{\textbf{k}',\textbf{k}}\bigl \rangle_{y}+\bigr \langle U_{\textbf{k},\textbf{k}'}U_{\textbf{k}',\textbf{k}}\bigl \rangle_{z}
\end{equation}
angular Brackets $ \bigr \langle...\bigl \rangle$ denotes the impurity average.
For elastic scattering:
\begin{equation}
    U_{\textbf{k},\textbf{k}'} = \frac{1}{2}\sum_{i}u^{i}_{0}e^{(\textbf{k}'-\textbf{k})\cdot\textbf{R}_{i}}\left[a^{2}e^{i(\phi-\phi')}+ b^{2}e^{-i(\phi-\phi')}+1\right]
    \end{equation}
\begin{equation}
     U_{\textbf{k'},\textbf{k}}=( U_{\textbf{k},\textbf{k}'})^{*}= \frac{1}{2}\sum_{j}u^{j}_{0}e^{-(\textbf{k}'-\textbf{k})\cdot\textbf{R}_{j}}\left[a^{2}e^{-i(\phi-\phi')}+ b^{2}e^{i(\phi-\phi')}+1\right],
\end{equation}
where $^*$ denotes the complex conjugation.
\begin{align}
   &U_{\textbf{k},\textbf{k}'}  U_{\textbf{k}',\textbf{k}} \nonumber\\&=
   \frac{1}{4}\sum_{i}\sum_{j}u^{i}_{0}u^{j}_{0}e^{-(\textbf{k}'-\textbf{k})\cdot(\textbf{R}_{i}-\textbf{R}_{j})}\left[a^{2}e^{i(\phi-\phi')}+ b^{2}e^{-i(\phi-\phi')}+1\right] \left[a^{2}e^{-i(\phi-\phi')}+ b^{2}e^{i(\phi-\phi')}+1\right] \nonumber \\
   &=
   \frac{1}{4}\sum_{i}\sum_{j}u^{i}_{0}u^{j}_{0}e^{-(\textbf{k}'-\textbf{k})\cdot(\textbf{R}_{i}-\textbf{R}_{j})} [a^{4}+a^{2}b^{2}e^{2i(\phi-\phi')}+a^{2}e^{i(\phi-\phi')}+a^{2}b^{2}e^{-2i(\phi-\phi')}+b^{4}\nonumber\\ &+b^{2}e^{-i(\phi-\phi')}+a^{2}e^{-i(\phi-\phi')}+b^{2}e^{i(\phi-\phi')}+1]\nonumber\\
 &  \bigr \langle U_{\textbf{k},\textbf{k}'}U_{\textbf{k}',\textbf{k}}\bigl \rangle_{e}=\frac{n_{0}u^{2}_{o}}{2}\left[\frac{(a^{4}+b^{4}+1)}{2}+\cos(\phi-\phi')+a^{2}b^{2}\cos2(\phi-\phi')\right]
\end{align}
where $n_{0}$ is the impurity concentration and $u_{o}$ is the average elastic impurity strength.
 {For components of magnetic scattering}
\begin{align}
     &\bigr \langle U_{\textbf{k},\textbf{k}'}U_{\textbf{k}',\textbf{k}}\bigl \rangle_{x}=\frac{n_{m}u^{2}_{x}}{4} \left[a(e^{i\phi}+e^{-i\phi'})+b(e^{i\phi'}+e^{-i\phi})\right]\left[a(e^{-i\phi}+e^{i\phi'})+b(e^{-i\phi'}+e^{i\phi})\right]\nonumber\\&
     = \frac{n_{m}u^{2}_{x}}{2}\left[1+\cos(\phi+\phi')+ab(2\cos(\phi-\phi')+\cos2\phi+\cos2\phi')\right] \nonumber\\
   &\bigr \langle U_{\textbf{k},\textbf{k}'}U_{\textbf{k}',\textbf{k}}\bigl \rangle_{y} = \frac{n_{m}u^{2}_{x}}{4}(i)\left[a(e^{-i\phi'}-e^{i\phi})+b(e^{-i\phi}-e^{i\phi'})\right]\left[a(e^{i\phi'}-e^{-i\phi})+b(e^{i\phi}-e^{-i\phi'})\right](-i)\nonumber \\&
    = \frac{n_{m}u^{2}_{x}}{2}\left[1-\cos(\phi+\phi')+ab(2\cos(\phi-\phi')-\cos2\phi'-\cos2\phi)\right]\nonumber\\&
\bigr\langle U_{\textbf{k},\textbf{k}'}U_{\textbf{k}',\textbf{k}}\bigl \rangle_{z}=\frac{n_{m}u^{2}_{z}}{4}\left[a^{2}e^{i(\phi-\phi')}-b^{2}e^{-i(\phi-\phi')}\right]\left[a^{2}e^{-i(\phi-\phi')}-b^{2}e^{i(\phi-\phi')}\right] \nonumber\\&
=\frac{n_{m}u^{2}_{z}}{4}\left[a^{4}+b^{4}-2a^{2}b^{2}\cos2(\phi-\phi')\right],
\end{align}
where $n_{m}$ denotes the concentration of magnetic impurities with  $u_{x,y,z}$   are their spatially averaged strengths.
\section{Greens's function and scattering rate}
The retarded (R) and advanced (A) Green's functions have the form
\begin{equation}
    G^{R/A}_{\textbf{k}}(\omega)=\frac{1}{\omega-\epsilon_{\textbf{k}}\pm i\frac{\hbar}{2\tau}}
\end{equation}
where the disorder-induced self-energy is described by the relaxation time $\tau$, which has the form
\begin{figure}[h]
    \centering
    \includegraphics[width=0.25\textwidth]{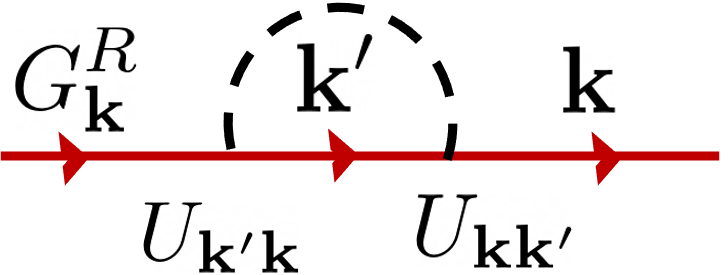}
    \caption{The retarded Green's function with its self-energy given by the first-order Born approximation. The solid and dash lines represent electron Green's function and impurity scattering.}
    \label{fig:gf}
\end{figure}
\begin{equation}
    \frac{1}{\tau}=\frac{2\pi}{\hbar}\sum_{\textbf{k}'}\bigr \langle U_{\textbf{k},\textbf{k}'}U_{\textbf{k}',\textbf{k}}\bigl \rangle_{imp}\delta(E_{F}-\epsilon_{\textbf{k}'})
\end{equation} 
\subsection{Elastic relaxation time}
\begin{align}
  &\frac{1}{\tau}_{e}= \frac{2\pi}{\hbar}\int_{0}^{\infty}\frac{k'dk'}{2\pi}\int_{0}^{2\pi}\frac{d\phi'}{2\pi}\bigr \langle U_{\textbf{k},\textbf{k}'}U_{\textbf{k}',\textbf{k}}\bigl \rangle_{e}\delta(E_{F}-\epsilon_{\textbf{k}'}) \nonumber\\ &
 = \frac{2\pi}{\hbar}\int_{0}^{\infty}\frac{k'dk'}{2\pi}\delta(E_{F}-\epsilon_{\textbf{k}'})\int_{0}^{2\pi}\frac{d\phi'}{2\pi}\bigr \langle U_{\textbf{k},\textbf{k}'}U_{\textbf{k}',\textbf{k}}\bigl \rangle_{e}\nonumber\\&
 =\frac{2\pi N_{F}}{\hbar}\int_{0}^{2\pi}\frac{n_{0}u^{2}_{o}}{2}\left[\frac{(a^{4}+b^{4}+1)}{2}+\cos(\phi-\phi')+a^{2}b^{2}\cos2(\phi-\phi')\right]\frac{d\phi'}{2\pi} 
\end{align}
\begin{equation}
   \frac{1}{\tau}_{e}=  \frac{2\pi N_{F}}{\hbar}\frac{n_{0}u^{2}_{o}}{4}(a^{4}+b^{4}+1).
\end{equation}
\subsection{Magnetic scattering time}
\begin{align}
      &\frac{1}{\tau_{m,x}}=\frac{2\pi}{\hbar}\sum_{\textbf{k}'}\bigr \langle U_{\textbf{k},\textbf{k}'}U_{\textbf{k}',\textbf{k}}\bigl \rangle_{x}\delta(E_{F}-\epsilon_{\textbf{k}'})\nonumber\\&
      = \frac{2\pi N_{F}}{h}\int_{0}^{2\pi}\frac{n_{m}u^{2}_{x}}{2}\left[1+\cos(\phi+\phi')+ab(2\cos(\phi-\phi')+\cos2\phi+\cos2\phi')\right]\frac{d\phi'}{2\pi} \nonumber\\
     &= \frac{2\pi N_{F}}{\hbar}\frac{n_{m}u^{2}_{x}}{2}(1+ab\cos2\phi)
\end{align}
\begin{equation}
   \frac{1}{\tau_{m,y}}= \frac{2\pi}{\hbar}\sum_{\textbf{k}'}\bigr \langle U_{\textbf{k},\textbf{k}'}U_{\textbf{k}',\textbf{k}}\bigl \rangle_{x}\delta(E_{F}-\epsilon_{\textbf{k}'})=\frac{2\pi N_{F}}{\hbar}\frac{n_{m}u^{2}_{y}}{2}(1-ab\cos2\phi) 
\end{equation}
\begin{equation}
     \frac{1}{\tau_{m,z}} =  \int_{0}^{2\pi}\frac{d\phi'}{2\pi}\bigr \langle U_{\textbf{k},\textbf{k}'}U_{\textbf{k}',\textbf{k}}\bigl \rangle_{z}= \frac{2\pi N_{F}}{\hbar}\frac{n_{m}u^{2}_{z}}{4}(a^{4}+b^{4}),
\end{equation}
where
\begin{equation}
     N_{F}= \int_{0}^{\infty}\frac{k'dk'}{2\pi}\delta(E_{F}-\epsilon_{\textbf{k}'}) =\frac{E_{F}}{2\pi(\hbar v_{F})^2}
\end{equation}
is the density of states in the two dimensions.
\subsection{Total relaxation time}
Total scattering time is the sum of elastic and magnetic scattering time,
\begin{align}
    &\frac{1}{\tau}= \frac{1}{\tau_{e}}+ \frac{1}{\tau_{m,x}}+ \frac{1}{\tau_{m,y}}+ \frac{1}{\tau_{m,z}} \nonumber\\&
     = \frac{2\pi N_{F}}{\hbar}\left[ \frac{n_{0}u^{2}_{o}}{4}(a^{4}+b^{4}+1)+\frac{n_{m}u^{2}_{x}}{2}(1+ab\cos2\phi)+\frac{n_{m}u^{2}_{y}}{2}(1-ab\cos2\phi) + \frac{n_{m}u^{2}_{z}}{4}(a^{4}+b^{4})\right] \nonumber\\&
     = \frac{\pi N_{F}}{\hbar}\left[ \frac{n_{0}u^{2}_{o}}{2}(a^{4}+b^{4}+1)+ n_{m}u^{2}_{x}(1+ab\cos2\phi)+n_{m}u^{2}_{y}(1-ab\cos2\phi) + \frac{n_{m}u^{2}_{z}}{2}(a^{4}+b^{4})\right].
   \end{align}
If in-plane isotropy is assumed$(u_{y}=u_{x})$,  then the total relaxation time becomes
\begin{equation}
    \frac{1}{\tau}=  \frac{2\pi N_{F}}{\hbar}\left[n_{0}u^{2}_{o}(a^{4}+b^{4}+1)+4n_{m}u^{2}_{x}+ n_{m}u^{2}_{z}(a^{4}+b^{4})\right],
\end{equation}
  from  the above equation, we can see that the total relaxation time is independent of polar angle $\phi$.

\section{Vertex correction to the velocity}
For pseudospin-1 fermions, the ladder diagram correction to velocity  should be considered. The diagram describes the iterative equation for the corrected velocity $\Tilde{v}^{x}_{\textbf{k}'}$,
\begin{figure}[h]
    \centering
    \includegraphics[width=0.35\textwidth]{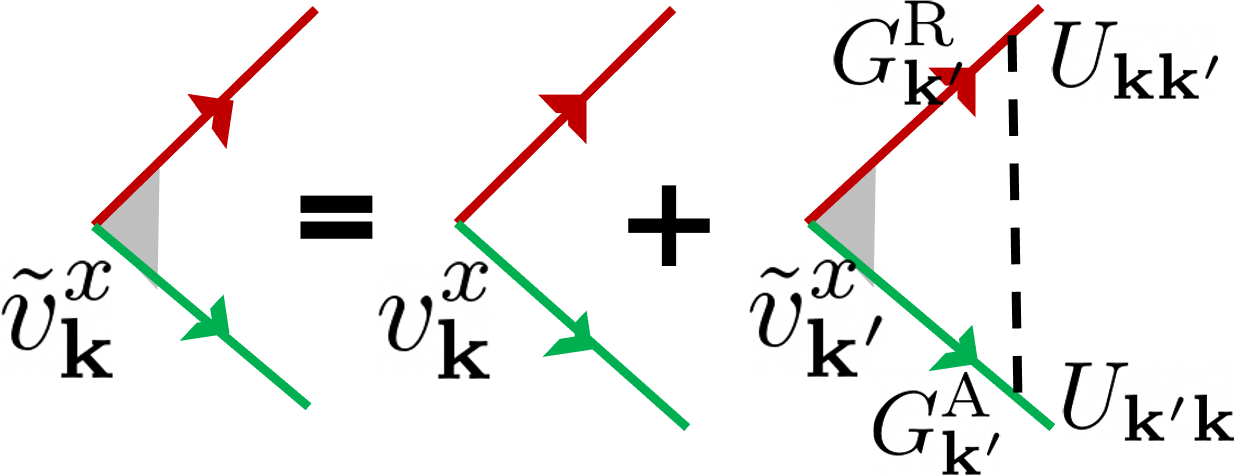}
    \caption{ The diagram for the vertex correction to the velocity, The arrow lines are for Green's functions. The dashed lines are for disorder scattering $(U)$.}
    \label{fig:3}
\end{figure}
\begin{equation}
\tilde{v}_{\mathbf{k}}^{i}=v_{\mathbf{k}}^{i}+\sum_{\mathbf{k}^{\prime}} G_{\mathbf{k}^{\prime}}^{\mathrm{R}} G_{\mathbf{k}^{\prime}}^{\mathrm{A}}\left\langle U_{\mathbf{k}, \mathbf{k}^{\prime}} U_{\mathbf{k}^{\prime}, \mathbf{k}}\right\rangle_{imp}\tilde{v}_{\mathbf{k}^{\prime}}^{i},
\label{Eq_vel_1}
\end{equation}
where $i=x/y$, In order to solve the above equation, a correction coefficient $\eta_{v}$ is introduced such that
\begin{equation}
   \tilde{v}_{\mathbf{k}}^{i}= \eta_{v}v_{\mathbf{k}}^{i},
\end{equation}
where the bare velocity $v_{\mathbf{k}}^{x}=v_{F}\cos\phi$, $v_{\mathbf{k}}^{y}=v_{F}\sin\phi$.
\begin{equation}
   \Tilde{v}^{x}_{\textbf{k}} = v_{\mathbf{k}}^{x}+\int_{0}^{\infty}\frac{k'dk'}{2\pi}G_{\mathbf{k}^{\prime}}^{\mathrm{R}} G_{\mathbf{k}^{\prime}}^{\mathrm{A}}\int_{0}^{2\pi}\frac{d\phi'}{2\pi}\eta_{v}v_{F}\cos\phi'\left\langle U_{\mathbf{k}, \mathbf{k}^{\prime}} U_{\mathbf{k}^{\prime}, \mathbf{k}}\right\rangle_{imp},
   \label{Eq_vel_2}
\end{equation}
\begin{multline}
    I  =\int_{0}^{\infty}\frac{k'dk'}{2\pi}G_{\mathbf{k}^{\prime}}^{\mathrm{R}} G_{\mathbf{k}^{\prime}}^{\mathrm{A}} \\
    = N_{F}\int_{-\infty}^{\infty}d\epsilon_\mathbf{{k}}\frac{1}{(\omega-\epsilon_{\textbf{k}}+ i\frac{\hbar}{2\tau})}\frac{1}{(\omega-\epsilon_{\textbf{k}}- i\frac{\hbar}{2\tau})},\\
\end{multline}
poles are given by
\begin{align}
\epsilon_\mathbf{{k}}= \left(\omega-\epsilon_{\textbf{k}}- i\frac{\hbar}{2\tau}\right), \  \left(\omega-\epsilon_{\textbf{k}}+ i\frac{\hbar}{2\tau}\right)   
\end{align}
above integral is solved by closing the contour in the upper half of the complex plane and using the residue theorem

{The residue at} $(\omega+ i\frac{\hbar}{2\tau})$ is   $\frac{\tau}{i\hbar}$. Finally above integration yields,
\begin{align}
    I = 2\pi i \times\frac{ N_{F}\tau}{i\hbar} 
    = \frac{2\pi  N_{F}\tau}{\hbar}.
\end{align}
Substituting the above results in Eq.~\ref{Eq_vel_2}, we obtain: 

\begin{equation}
   \Tilde{v}^{x}_{\textbf{k}} = v_{\mathbf{k}}^{i}+\frac{2\pi N_{F}\tau}{\hbar}\eta_{v}v_{F}  \int_{0}^{2\pi}\frac{d\phi'}{2\pi}\cos\phi'\left\langle U_{\mathbf{k}, \mathbf{k}^{\prime}} U_{\mathbf{k}^{\prime}, \mathbf{k}}\right\rangle_{imp},
   \label{Eq_vel_3}
\end{equation}
where we have
\begin{align}
     &\int_{0}^{2\pi}\frac{d\phi'}{2\pi}\cos\phi'\left\langle U_{\mathbf{k}, \mathbf{k}^{\prime}} U_{\mathbf{k}^{\prime}, \mathbf{k}}\right\rangle_{imp} =\nonumber\\ & \int_{0}^{2\pi}\frac{d\phi'}{2\pi}\cos\phi'\left[\left\langle U_{\mathbf{k}, \mathbf{k}^{\prime}} U_{\mathbf{k}^{\prime}, \mathbf{k}}\right\rangle_{e}+\left\langle U_{\mathbf{k}, \mathbf{k}^{\prime}} U_{\mathbf{k}^{\prime}, \mathbf{k}}\right\rangle_{x}+\left\langle U_{\mathbf{k}, \mathbf{k}^{\prime}} U_{\mathbf{k}^{\prime}, \mathbf{k}}\right\rangle_{y}+\left\langle U_{\mathbf{k}, \mathbf{k}^{\prime}} U_{\mathbf{k}^{\prime}, \mathbf{k}}\right\rangle_{z}\right]
\end{align}
\begin{align}
    &\int_{0}^{2\pi}\frac{d\phi'}{2\pi}\cos\phi'\left\langle U_{\mathbf{k}, \mathbf{k}^{\prime}} U_{\mathbf{k}^{\prime}, \mathbf{k}}\right\rangle_{e} =\nonumber\\&\frac{n_{0}u_{0}^{2}}{2} \int_{0}^{2\pi}\frac{d\phi'}{2\pi}\cos\phi'\left[\frac{(a^{4}+b^{4}+1)}{2}+\cos(\phi-\phi')+a^{2}b^{2}\cos2(\phi-\phi')\right]=\nonumber \\&
     \frac{n_{0}u_{0}^{2}}{2} \int_{0}^{2\pi}\frac{d\phi'}{2\pi}\cos\phi'\cos(\phi-\phi')
    = \frac{n_{0}u_{0}^{2}}{4}\cos\phi
\end{align}
\begin{align}
     &\int_{0}^{2\pi}\frac{d\phi'}{2\pi}\cos\phi'\left\langle U_{\mathbf{k}, \mathbf{k}^{\prime}} U_{\mathbf{k}^{\prime}, \mathbf{k}}\right\rangle_{x} =\nonumber\\& \frac{n_{m}u^{2}_{x}}{2}\int_{0}^{2\pi}\frac{d\phi'}{2\pi}\cos\phi' \left[1+\cos(\phi+\phi')+ab(2\cos(\phi-\phi')+\cos2\phi+\cos2\phi')\right]
    \end{align}
    \begin{equation}
         = \frac{n_{m}u^{2}_{x}}{4}(1+2ab)\cos\phi
    \end{equation}
    \begin{align}
     &\int_{0}^{2\pi}\frac{d\phi'}{2\pi}\cos\phi'\left\langle U_{\mathbf{k}, \mathbf{k}^{\prime}} U_{\mathbf{k}^{\prime}, \mathbf{k}}\right\rangle_{y} =\nonumber\\ &\frac{n_{m}u^{2}_{y}}{2}\int_{0}^{2\pi}\frac{d\phi'}{2\pi}\cos\phi' \left[1-\cos(\phi+\phi')+ab(2\cos(\phi-\phi')-\cos2\phi-\cos2\phi')\right]
    \end{align}
    \begin{equation}
         = -\frac{n_{m}u^{2}_{y}}{4}(1-2ab)\cos\phi
    \end{equation}
    \begin{equation}
         \int_{0}^{2\pi}\frac{d\phi'}{2\pi}\cos\phi'\left\langle U_{\mathbf{k}, \mathbf{k}^{\prime}} U_{\mathbf{k}^{\prime}, \mathbf{k}}\right\rangle_{z}= 0, 
    \end{equation}
    now Eq.~\ref{Eq_vel_3}  becomes
    \begin{equation}
        \eta_{v}=1+\frac{2\pi N_{F}\tau}{\hbar}\eta_{v}\left[\frac{n_{0}u_{0}^{2}}{4}+\frac{n_{m}u^{2}_{x}}{4}(1+2ab)\ -\frac{n_{m}u^{2}_{y}}{4}(1-2ab)\right]
    \end{equation} 
    where we have,
  \begin{align}
  &\frac{n_{0}u_{0}^{2}}{4}=\frac{\hbar}{2\pi N_{F}\tau_{e}(a^{4}+b^{4}+1)},\nonumber \\
  &\frac{n_{m}u^{2}_{x}}{4}=  \frac{\hbar}{4\pi N_{F}\tau_{m,x}(1+ab\cos2\phi)},\nonumber \\
  &\frac{n_{m}u^{2}_{y}}{4}= \frac{\hbar}{4\pi N_{F}\tau_{m,y}(1-ab\cos2\phi)}.
\end{align}  
Using the above relations, we get,
\begin{equation}
    \eta_{v}=1+\eta_{v}\left[\frac{\tau}{\tau_{e}(a^{4}+b^{4}+1)}+\frac{\tau(1+2ab)}{2\tau_{m,x}(1+ab\cos2\phi)}-\frac{\tau(1-2ab)}{2\tau_{m,y}(1-ab\cos2\phi)}\right],
\end{equation}
finally, the vertex correction factor is given by,
\begin{equation}
    \eta_{v}=\frac{1}{1-\left[\frac{\tau}{\tau_{e}(a^{4}+b^{4}+1)}+\frac{\tau(1+2ab)}{2\tau_{m,x}(1+ab\cos2\phi)}-\frac{\tau(1-2ab)}{2\tau_{m,y}(1-ab\cos2\phi)}\right]}.
\end{equation}
If we assume in-plane isotropy $u_{x}= u_{y}$, then 
\begin{align}
    &\eta_{v}=1+\frac{2\pi N_{F}\tau}{\hbar}\eta_{v}\left[\frac{n_{0}u_{0}^{2}}{4}+n_{m}u^{2}_{x}(ab)\right] \nonumber\\&
   \eta_{v}=1+\eta_{v}\left[\frac{\tau}{\tau_{e}(a^{4}+b^{4}+1)}+(2ab)\frac{\tau}{\tau_{m,x}}\right]\nonumber\\&
    \eta_{v}=\frac{1}{1-\left[\frac{\tau}{\tau_{e}(a^{4}+b^{4}+1)}+(2ab)\frac{\tau}{\tau_{m,x}}\right]}.  
   \end{align}
\section{Bare and dressed Hikami boxes}
The quantum interference correction to conductivity is obtained by the calculation of bare Hiakmi box and two dressed Hikami boxes.
\subsection{Bare Hikami box}
The bare Hikami box at zero temperature is calculated as
\begin{equation}
     \sigma_{0}^{F}=\frac{e^{2}\hbar}{2\pi}\sum_{\textbf{q}}\Gamma(\textbf{q})\sum_{\textbf{k}}\Tilde v_{\textbf{k}}^{x}\Tilde v^{x}_{\textbf{q-k}}G_{\mathbf{k}}^{\mathrm{R}} G_{\mathbf{k}}^{\mathrm{A}}G_{\mathbf{q-k}}^{\mathrm{R}} G_{\mathbf{q-k}}^{\mathrm{A}},
\end{equation}
\begin{figure}[h]
    \centering
    \includegraphics[width=0.6\textwidth]{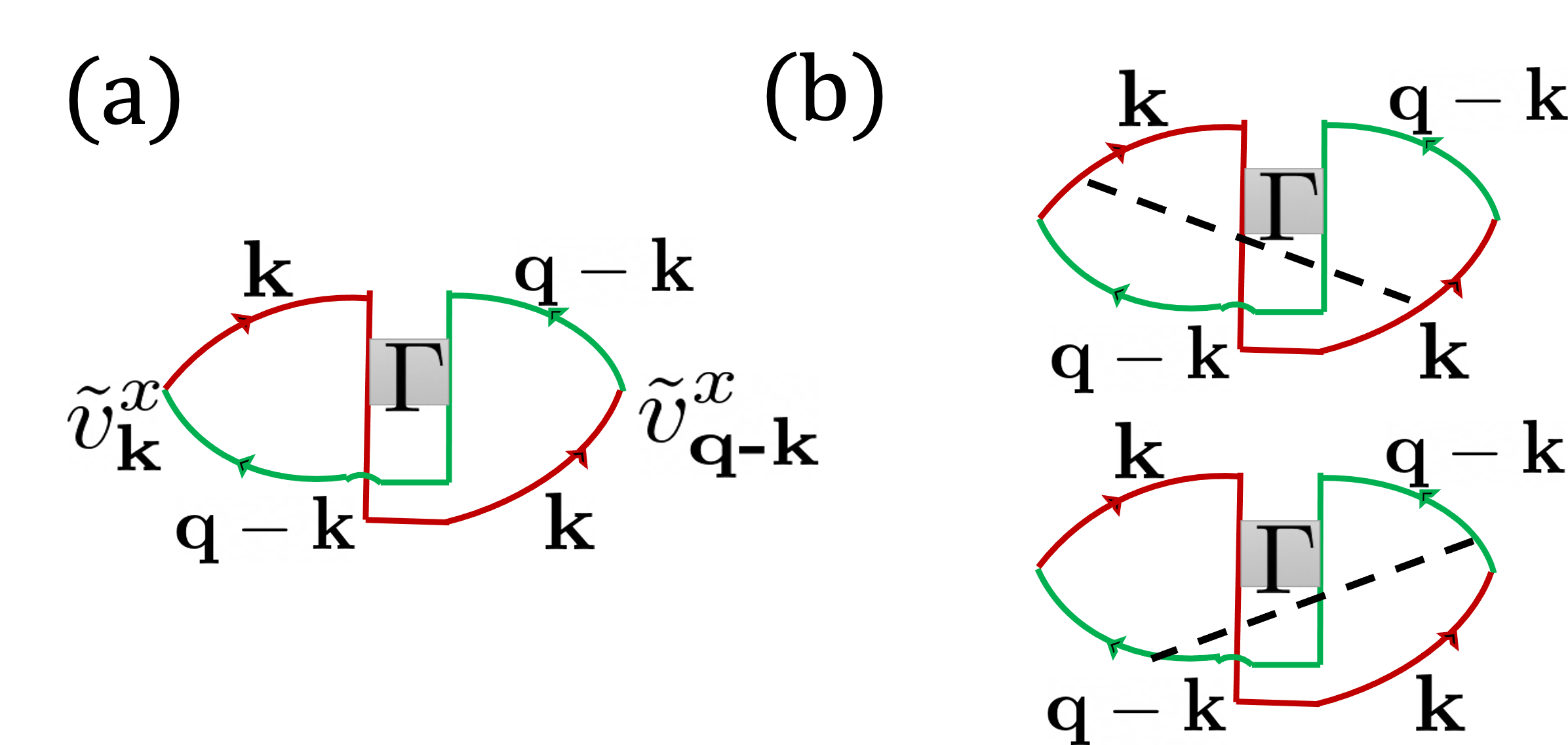}
    \caption{The diagrams for the quantum interference
correction to conductivity of pseudospin-1 fermions. The arrowed
solid and dashed lines represent the Green’s functions and
impurity scattering, respectively. (a) The bare and
(b) two dressed Hikami boxes give the quantum conductivity
correction from the maximally crossed diagrams.}
    \label{fig:5}
\end{figure}
retarded and advanced Green's function is given by 
\begin{align}
     G^{R/A}_{\mathbf{k}}(\omega)=\frac{1}{\omega-\epsilon_{\textbf{k}}\pm i\frac{\hbar}{2\tau}}, \    
      G^{R/A}_{\mathbf{q-k}}(\omega)=\frac{1}{\omega-\epsilon_{\mathbf{q-k}}\pm i\frac{\hbar}{2\tau}}
\end{align}
In the small $\mathbf{q}$ limit,
\begin{equation}
 \epsilon_{\mathbf{q-k}}\approx \epsilon_{\textbf{k}}- \hbar\mathbf{v_{F}.q},
\end{equation}
$\Gamma(\textbf{q})$ is the vertex function which totally depends on $\mathbf{q}$(incoming momentum). The contribution for small $\mathbf{q}$ dominates the summation, approximately we can take $\mathbf{k}$ summation for  $\mathbf{q}=0$  first.
\begin{align}
    \sigma_{0}^{F}=\frac{e^{2}\hbar \eta^{2}_{v} v^{2}_{F}}{2\pi}\sum_{\textbf{q}}\Gamma(\textbf{q})\sum_{\textbf{k}}\cos\phi\cos(\phi-\theta)G_{\mathbf{k}}^{\mathrm{R}} G_{\mathbf{k}}^{\mathrm{A}}G_{\mathbf{q-k}}^{\mathrm{R}} G_{\mathbf{q-k}}^{\mathrm{A}} \\
   =\frac{e^{2}\hbar \eta^{2}_{v} v^{2}_{F}}{2\pi}\int_{0}^{\infty}\frac{kdk}{2\pi}G_{\mathbf{k}}^{\mathrm{R}} G_{\mathbf{k}}^{\mathrm{A}}G_{\mathbf{q-k}}^{\mathrm{R}} G_{\mathbf{q-k}}^{\mathrm{A}}\int_{0}^{2\pi}\frac{d\phi}{2\pi}\cos\phi\cos(\phi-\theta).
\end{align}
Let's assume that
\begin{align}
    I= \int_{0}^{\infty}\frac{kdk}{2\pi}G_{\mathbf{k}}^{\mathrm{R}} G_{\mathbf{k}}^{\mathrm{A}}G_{\mathbf{q-k}}^{\mathrm{R}} G_{\mathbf{q-k}}^{\mathrm{A}}, \\
    = N_{F}\int_{-\infty}^{\infty}d\epsilon_{\mathbf{k}}G_{\mathbf{k}}^{\mathrm{R}} G_{\mathbf{k}}^{\mathrm{A}}G_{\mathbf{q-k}}^{\mathrm{R}} G_{\mathbf{q-k}}^{\mathrm{A}}
   \end{align}
   \begin{align}
       = N_{F}\int_{-\infty}^{\infty}d\epsilon_{\mathbf{k}} \frac{1}{(\omega-\epsilon_{\mathbf{k}}+\frac{i\hbar}{2\tau})} \frac{1}{(\omega-\epsilon_{\mathbf{k}}-\frac{i\hbar}{2\tau})}\frac{1}{(\omega-\epsilon_{\mathbf{k}}+\hbar \mathbf{v_{F}.q}+\frac{i\hbar}{2\tau})}\frac{1}{(\omega-\epsilon_{\mathbf{k}}+\hbar \mathbf{v_{F}.q}-\frac{i\hbar}{2\tau})}
   \end{align}
using the technique of contour integration, poles are
\begin{align}
 \epsilon_{\mathbf{k}}=  \omega+\frac{i\hbar}{2\tau} ;\   \omega-\frac{i\hbar}{2\tau}; \ \omega+\hbar \mathbf{v_{F}\cdot q}+\frac{i\hbar}{2\tau};\ \omega+\hbar \mathbf{v_{F}\cdot q}-\frac{i\hbar}{2\tau}
\end{align}
contour is closed in the upper half of the complex plane, and the sum of residues is given by,
\begin{itemize}
    \item \textbf{residue at} $\epsilon_{\mathbf{k}}=  \omega+\frac{i\hbar}{2\tau}$
\end{itemize}
\begin{equation}
    \frac{1}{\frac{i\hbar}{\tau}(-\hbar \mathbf{v_{F}.q})(\frac{i\hbar}{\tau}-\hbar \mathbf{v_{F}\cdot q})}
\end{equation}
\begin{itemize}
    \item \textbf{residue at} $\epsilon_{\mathbf{k}}=  \omega+\frac{i\hbar}{2\tau}+\hbar \mathbf{v_{F} \cdot q}$,
\end{itemize}
\begin{equation}
    \frac{1}{\frac{i\hbar}{\tau}(\hbar \mathbf{v_{F} \cdot q})(\frac{i\hbar}{\tau}+\hbar \mathbf{v_{F} \cdot q})},
\end{equation}
the sum of residues is 
\begin{equation}
    \frac{1}{\frac{i\hbar}{\tau}((\hbar \mathbf{v_{F}\cdot q})^{2}+\frac{\hbar^{2}}{\tau^{2}})},
\end{equation}
in the limit of $\mathbf{q}\to 0$, sum of residues becomes
\begin{equation}
    \frac{-2\tau^{3}}{i\hbar^{3}},
\end{equation}
finally, the value of  the integral is
\begin{equation}
    I= \frac{-4\pi N_{F}\tau^{3}}{\hbar^{3}}.
\end{equation}
Finally, we obtain:
\begin{equation}
     \sigma_{0}^{F}=\frac{-2e^{2}N_{F}\tau^{3}\eta^{2}_{v}v^{2}_{F}}{\hbar^{2}}\sum_{\textbf{q}}\Gamma(\textbf{q})\int_{0}^{2\pi}\frac{d\phi}{2\pi}[\cos^{2}\phi\cos\theta+\cos\phi\sin\phi\sin\theta]
\end{equation}
as $\mathbf{q}\to 0$, $\theta \to 0$ $\to$ $\cos\theta \to 1$. Finally, the conductivity correction at zero temperature from  one bare Hikami box is 
\begin{equation}
    \sigma_{0}^{F}=\frac{-e^{2}N_{F}\tau^{3}\eta^{2}_{v}v^{2}_{F}}{\hbar^{2}}\sum_{\textbf{q}}\Gamma(\textbf{q})
\end{equation}
\subsection{ Two dressed Hikami boxes}
Two dressed Hikami boxes denoted as $\sigma_{R}^{F}$ and $\sigma_{A}^{F}$
\begin{multline}
    \sigma_{R}^{F}= \frac{e^{2}\hbar}{2\pi}\sum_{\textbf{q}}\Gamma(\textbf{q})\sum_{\textbf{k}}\sum_{\textbf{k}_{1}}\Tilde v_{\textbf{k}}^{x}\Tilde v^{x}_{\textbf{q}-\textbf{k}_{1}}G_{\mathbf{k}}^{\mathrm{R}}G_{\mathbf{k_1}}^{\mathrm{R}}G_{\mathbf{q-k}}^{\mathrm{R}} G_{\mathbf{q-k_1}}^{\mathrm{R}}G_{\mathbf{k}}^{\mathrm{A}} G_{\mathbf{q-k_1}}^{\mathrm{A}}\langle U_{\mathbf{k}_{1},\textbf{k}}U_{\mathbf{q-k_{1}},\mathbf{q-k}}\bigl \rangle_{imp},
\end{multline}
contribution from small $\mathbf{q}$ dominates the summation, approximately we can evaluate $\mathbf{k}$ integrals for $\textbf{q=0}$
\begin{align}
  &\sigma_{R}^{F}= \frac{e^{2}\hbar}{2\pi}\sum_{\textbf{q}}\Gamma(\textbf{q})\sum_{\textbf{k}}\Tilde v_{\textbf{k}}^{x} G_{\mathbf{k}}^{\mathrm{R}} G_{\mathbf{k}}^{\mathrm{A}}G_{\mathbf{q-k}}^{\mathrm{R}}\sum_{\textbf{k}_{1}}\Tilde v^{x}_{\textbf{q}-\textbf{k}_{1}}G_{\mathbf{k_1}}^{\mathrm{R}}G_{\mathbf{q-k_1}}^{\mathrm{R}} G_{\mathbf{q-k_1}}^{\mathrm{A}}\langle U_{\mathbf{k}_{1},\textbf{k}}U_{\mathbf{q-k_{1}},\mathbf{q-k}}\bigl \rangle_{imp}\nonumber\\
  &= \frac{e^{2}\hbar}{2\pi}\sum_{\textbf{q}}\Gamma(\textbf{q})\int_{0}^{\infty}\frac{kdk}{2\pi}G_{\mathbf{k}}^{\mathrm{R}} G_{\mathbf{k}}^{\mathrm{A}}G_{\mathbf{q-k}}^{\mathrm{R}}\int_{0}^{\infty}\frac{k_{1}dk_{1}}{2\pi}G_{\mathbf{k_1}}^{\mathrm{R}}G_{\mathbf{q-k_1}}^{\mathrm{R}} G_{\mathbf{q-k_1}}^{\mathrm{A}}\nonumber \\ &\times\int_{0}^{2\pi}\frac{d\phi}{2\pi}\Tilde v_{\textbf{k}}^{x}\int_{0}^{2\pi}\frac{d\phi_{1}}{2\pi}\Tilde v_{\mathbf{q-k_{1}}}^{x}\langle U_{\mathbf{k}_{1},\textbf{k}}U_{\mathbf{q-k_{1}},\mathbf{q-k}}\bigl \rangle_{imp}
\end{align}
Solution to the $\mathbf{k}$-integral is:
\begin{align}
     &I = \int_{0}^{\infty}\frac{kdk}{2\pi}G_{\mathbf{k}}^{\mathrm{R}} G_{\mathbf{k}}^{\mathrm{A}}G_{\mathbf{q-k}}^{\mathrm{R}} \nonumber \\
     &= N_{F}\int_{-\infty}^{\infty}d\epsilon_{\mathbf{k}} \frac{1}{(\omega-\epsilon_{\mathbf{k}}+\frac{i\hbar}{2\tau})}\frac{1}{(\omega-\epsilon_{\mathbf{k}}-\frac{i\hbar}{2\tau})}\frac{1}{(\omega-\epsilon_{\mathbf{k}}+\hbar\mathbf{v_{F} \cdot q}+\frac{i\hbar}{2\tau})}
     \end{align}
The poles are given by
\begin{align}
  \epsilon_{\mathbf{k}}= \omega+ \frac{i\hbar}{2\tau},\ \omega- \frac{i\hbar}{2\tau}, \ \omega+\hbar\mathbf{v_{F}\cdot q}+\frac{i\hbar}{2\tau}.
\end{align}
The contour is closed in the upper half of the complex plane. The residues are: 
\begin{itemize}
    \item \textbf{Residue at} $\omega+ \frac{i\hbar}{2\tau}$
    \end{itemize}
  \begin{equation}
      \frac{1}{(-\hbar\mathbf{v_{F}\cdot q})\frac{i\hbar}{\tau}}
  \end{equation}
\begin{itemize}
    \item \textbf{Residue at} $\omega+ \frac{i\hbar}{2\tau}+\hbar\mathbf{v_{F}\cdot q}$
    \end{itemize}
    \begin{equation}
        \frac{1}{\hbar\mathbf{v_{F}\cdot q}(\hbar\mathbf{v_{F} \cdot q}+\frac{i\hbar}{\tau})}
    \end{equation}
  \begin{itemize}
    \item \textbf{ Sum of Residues } 
    \end{itemize}  
     \begin{equation}
        \frac{-1}{\frac{i\hbar}{\tau}(\hbar\mathbf{v_{F}\cdot q}+\frac{i\hbar}{\tau})}
    \end{equation}
    as $\mathbf{q}\to 0$, the value of integral $I$ becomes
    \begin{equation}
        I= -\frac{2\pi i N_{F}\tau^{2}}{\hbar^{2}}.
    \end{equation}
\begin{align}
   &I_{1}= \int_{0}^{\infty}\frac{k_{1}dk_{1}}{2\pi}G_{\mathbf{k_1}}^{\mathrm{R}}G_{\mathbf{q-k_1}}^{\mathrm{R}} G_{\mathbf{q-k_1}}^{\mathrm{A}} \nonumber\\
        &=   N_{F}\int_{-\infty}^{\infty}d\epsilon_{\mathbf{k}} \frac{1}{(\omega-\epsilon_{\mathbf{k}}+\frac{i\hbar}{2\tau})}\frac{1}{(\omega-\epsilon_{\mathbf{k_{1}}}+\hbar\mathbf{v_{F}\cdot q}-\frac{i\hbar}{2\tau})}\frac{1}{(\omega-\epsilon_{\mathbf{k_{1}}}+\hbar\mathbf{v_{F}\cdot q}-\frac{i\hbar}{2\tau})}\nonumber\\
        &= -\frac{2\pi i N_{F}\tau^{2}}{\hbar^{2}}
\end{align}
 now Eq. (64) becomes
 \begin{align}
   &\sigma_{R}^{F}=-\frac{2\pi e^{2}N^{2}_{F}\tau^{4}\eta^{2}_{v}v^{2}_{F}}{\hbar^{3}}\sum_{\textbf{q}}\Gamma(\textbf{q})\int_{0}^{2\pi}\frac{d\phi}{2\pi}\cos\phi\int_{0}^{2\pi}\frac{d\phi_{1}}{2\pi}\cos\phi_{1}\langle U_{\mathbf{k}_{1},\textbf{k}}U_{\mathbf{-k_{1}},\mathbf{-k}}\bigl \rangle_{imp} \nonumber\\
   &=-\frac{2\pi e^{2}N^{2}_{F}\tau^{4}\eta^{2}_{v}v^{2}_{F}}{\hbar^{3}}\sum_{\textbf{q}}\Gamma(\textbf{q})\int_{0}^{2\pi}\frac{d\phi}{2\pi}\cos\phi\int_{0}^{2\pi}\frac{d\phi_{1}}{2\pi}\cos\phi_{1}\frac{\hbar}{2\pi N_{F}\tau}\sum_{m=-2}^{2}\sum_{n=-2}^{2}z_{mn}e^{im\phi_{1}}e^{in\phi},\nonumber\\
   &= -\frac{ e^{2}N_{F}\tau^{3}\eta^{2}_{v}v^{2}_{F}}{\hbar^{2}}\sum_{\textbf{q}}\Gamma(\textbf{q})\int_{0}^{2\pi}\frac{d\phi}{2\pi}\cos\phi\int_{0}^{2\pi}\frac{d\phi_{1}}{2\pi}\cos\phi_{1}\sum_{m=-2}^{2}\sum_{n=-2}^{2}z_{mn}e^{im\phi_{1}}e^{in\phi},
   \end{align}
Integral over $\phi$ and $\phi_{1}$ survives only for $m,n =\pm 1$, therefore
\begin{multline}
 = -\frac{ e^{2}N_{F}\tau^{3}\eta^{2}_{v}v^{2}_{F}}{\hbar^{2}}\sum_{\textbf{q}}\Gamma(\textbf{q})\frac{[z_{11}+z_{1-1}+z_{-11}+z_{-1-1}]}{4} \\
 -\frac{ e^{2}N_{F}\tau^{3}\eta^{2}_{v}v^{2}_{F}}{\hbar^{2}}\eta_{H}\sum_{\textbf{q}}\Gamma(\textbf{q})
\end{multline}
   
\begin{multline}
     \sigma_{A}^{F}= \frac{e^{2}\hbar}{2\pi}\sum_{\textbf{q}}\Gamma(\textbf{q})\sum_{\textbf{k}}\sum_{\textbf{k}_{1}}\Tilde v_{\textbf{k}}^{x}\Tilde v^{x}_{\textbf{q}-\textbf{k}_{1}}G_{\mathbf{k}}^{\mathrm{R}}G_{\mathbf{q-k_1}}^{\mathrm{R}}G_{\mathbf{k}}^{\mathrm{A}}G_{\mathbf{k_1}}^{\mathrm{A}}G_{\mathbf{q-k}}^{\mathrm{A}} G_{\mathbf{q-k_1}}^{\mathrm{A}}\langle U_{\textbf{k},\mathbf{k}_{1}}U_{\mathbf{q-k},\mathbf{q-k_1}}\bigl \rangle_{imp},
\end{multline}
\begin{align}
     \sigma_{A}^{F}= \sigma_{R}^{F}=  -\frac{ e^{2}N_{F}\tau^{3}\eta^{2}_{v}v^{2}_{F}}{\hbar^{2}}\eta_{H}\sum_{\textbf{q}}\Gamma(\textbf{q})
\end{align}
with 
\begin{equation}
    \eta_{H}=-\frac{1}{2}\left(1-\eta^{-1}_{v}\right),
\end{equation}
where 
\begin{equation}
     \eta_{v}=\frac{1}{1-\left[\frac{\tau}{\tau_{e}(a^{4}+b^{4}+1)}+\frac{\tau(1+2ab)}{2\tau_{m,x}(1+ab\cos2\phi)}-\frac{\tau(1-2ab)}{2\tau_{m,y}(1-ab\cos2\phi)}\right]}.
\end{equation}
If in-plane isotropy is assumed, then 
 \begin{equation}
    \eta_{v}=\frac{1}{1-\left[\frac{\tau}{\tau_{e}(a^{4}+b^{4}+1)}+2ab\frac{\tau}{\tau_{m,x}}\right]}.  
   \end{equation}
The total contribution is given by the summation of one bare and two dressed Hikami boxes
\begin{equation*}
       \sigma^{F}= \sigma_{0}^{F}+ \sigma_{R}^{F} + \sigma_{A}^{F} 
          \end{equation*}
          \begin{equation}
                = -\frac{ e^{2}N_{F}\tau^{3}\eta^{2}_{v}v^{2}_{F}}{\hbar^{2}}(1+2\eta_{H})\sum_{\textbf{q}}\Gamma(\textbf{q}).  
          \end{equation}
\section{Bethe-Salpeter equation}
Now the goal is to derive the vertex function $\Gamma(\textbf{q})$ for the maximally crossed diagrams. First we need to evaluate the bare vertex $\Gamma^{0}(\textbf{q})$, which has the form in the limit of $\textbf{q} \to 0 $
\begin{multline}
     \Gamma^{0}_{\mathbf{k_{1},k_{2}}}\equiv \langle U_{\mathbf{k_1},\mathbf{k}_{2}}U_{\mathbf{-k_1},\mathbf{-k_2}}\bigl \rangle_{imp} \\
     = \langle U_{\mathbf{k_1},\mathbf{k}_{2}}U_{\mathbf{-k_1},\mathbf{-k_2}}\bigl \rangle_{e}+\langle U_{\mathbf{k_1},\mathbf{k}_{2}}U_{\mathbf{-k_1},\mathbf{-k_2}}\bigl \rangle_{x}+\langle U_{\mathbf{k_1},\mathbf{k}_{2}}U_{\mathbf{-k_1},\mathbf{-k_2}}\bigl \rangle_{y}+\langle U_{\mathbf{k_1},\mathbf{k}_{2}}U_{\mathbf{-k_1},\mathbf{-k_2}}\bigl \rangle_{z},
\end{multline}
all the $5\times5$ matrices involved in the calculation of vertex function is written as
\begin{equation}
    \begin{pmatrix}
        a_{-2-2} &  a_{-2-1} &  a_{-20} &  a_{-21} & a_{-22} \\
          a_{-1-2} &  a_{-1-1} &  a_{-10} &  a_{-11} & a_{-12} \\
            a_{0-2} &  a_{0-1} &  a_{00} &  a_{01} & a_{02} \\
              a_{1-2} &  a_{1-1} &  a_{10} &  a_{11} & a_{12} \\
               a_{2-2} &  a_{2-1} &  a_{20} &  a_{21} & a_{22}
          \end{pmatrix}
\end{equation}
\subsection{For the elastic scattering}
\begin{equation}
    U_{\mathbf{k_{1},k_{2}}}= \frac{1}{2}\sum_{i}u^{i}_{0}e^{(\mathbf{k'-k}).R_{i}}\left[a^{2}e^{i(\phi-\phi')}+ b^{2}e^{-i(\phi-\phi')}+1 \right],
\end{equation}
as $\mathbf{k}\to \mathbf{-k}$, $\phi \to \pi+\phi$, the difference between polar angles $(\phi-\phi')$ remains same. Therefore 
\begin{equation}
  U_{\mathbf{-k_{1}, -k_{2}}}= \frac{1}{2}\sum_{j}u^{j}_{0}e^{-(\mathbf{k'-k}).R_{j}}\left[a^{2}e^{i(\phi-\phi')}+ b^{2}e^{-i(\phi-\phi')}+1 \right],   
\end{equation}
\begin{align}
    &\langle U_{\mathbf{k_1},\mathbf{k}_{2}}U_{\mathbf{-k_1},\mathbf{-k_2}}\bigl \rangle_{e}\nonumber\\&= \frac{n_{0}u_{0}^{2}}{4} \left[a^{2}e^{i(\phi-\phi')}+ b^{2}e^{-i(\phi-\phi')}+1 \right] \left[a^{2}e^{i(\phi-\phi')}+ b^{2}e^{-i(\phi-\phi')}+1 \right] \nonumber\\
    &= \frac{n_{0}u_{0}^{2}}{4}\left[a^{4}e^{2i(\phi-\phi')}+ 2a^{2}b^{2}+1 + 2a^{2}e^{i(\phi-\phi')}+b^{4}e^{-2i(\phi-\phi')}+2b^{2}e^{-i(\phi-\phi')}\right]
  \end{align}
where we have
\begin{align}
    \frac{n_{0}u_{0}^{2}}{4}=  \frac{\hbar}{2\pi N_{F}\tau}\frac{\tau}{\tau_{e}(a^{4}+b^{4}+1)} \\
    \alpha_{e}= \frac{\tau}{\tau_{e}},
\end{align}
finally the bare vertex for elastic scattering is given as
\begin{equation}
     \langle U_{\mathbf{k_1},\mathbf{k}_{2}}U_{\mathbf{-k_1},\mathbf{-k_2}}\bigl \rangle_{e}=\frac{\hbar}{2\pi N_{F}\tau}\sum_{m=-2}^{2}\sum_{n=-2}^{2}z^{e}_{mn}e^{im\phi_{1}}e^{in\phi_{2}}
\end{equation}
where
\begin{equation}
 z^{e}_{mn} =   \begin{pmatrix}
 0 & 0 & 0 & 0 & \alpha_{e}\frac{b^{4}}{(a^{4}+b^{4}+1)}\\
        0 & 0 & 0 & \alpha_{e}\frac{2b^{2}}{(a^{4}+b^{4}+1)} & 0 \\
        0 & 0 & \alpha_{e}\frac{(2a^{2}b^{2}+1)}{(a^{4}+b^{4}+1)} & 0 & 0 \\
        0 &\alpha_{e}\frac{2a^{2}}{(a^{4}+b^{4}+1)} & 0 & 0 & 0\\
        \alpha_{e}\frac{a^{4}}{(a^{4}+b^{4}+1)} & 0 & 0 & 0 & 0
         \end{pmatrix}
\end{equation}
\subsection{For the magnetic impurity}
\begin{itemize}
    \item \textbf{x component}
\end{itemize}
\begin{equation}
    U_{\mathbf{k_1},\mathbf{k}_{2}}= \frac{1}{2}\sum_{i}u^{i}_{x}e^{(\mathbf{k'-k}).\mathbf{R}_{i}}\left[a(e^{i\phi_{1}}+e^{-i\phi_{2}})+b(e^{i\phi_{2}}+e^{-i\phi_{1}})\right]
\end{equation}
as  $\mathbf{k}\to \mathbf{-k}$, $\phi \to \pi+\phi$, $e^{i\phi}\to -e^{i\phi}$, 
\begin{equation}
     U_{\mathbf{k_1},\mathbf{k}_{2}}= -\frac{1}{2}\sum_{j}u^{j}_{x}e^{-(\mathbf{k'-k}).\mathbf{R}_{i}}\left[a(e^{i\phi_{1}}+e^{-i\phi_{2}})+b(e^{i\phi_{2}}+e^{-i\phi_{1}})\right]
\end{equation}
\begin{align}
    &\langle U_{\mathbf{k_1},\mathbf{k}_{2}}U_{\mathbf{-k_1},\mathbf{-k_2}}\bigl \rangle_{x}= -\frac{n_{m}u^{2}_{x}}{4}\left[a(e^{i\phi_{1}}+e^{-i\phi_{2}})+b(e^{i\phi_{2}}+e^{-i\phi_{1}})\right]^{2}\nonumber\\&
    = -\frac{n_{m}u^{2}_{x}}{4}[a^{2}e^{2i\phi_{1}}+a^{2}e^{-2i\phi_{2}}+2a^{2}e^{i\phi_{1}}e^{-i\phi_{2}}+b^{2}e^{2i\phi_{2}}+b^{2}e^{-2i\phi_{1}}+2b^{2}e^{-i\phi_{1}}e^{i\phi_{2}}+2abe^{i\phi_{1}}e^{i\phi_{2}}\nonumber\\&
    +4ab+2abe^{-i\phi_{1}}e^{-i\phi_{2}}], \nonumber\\&
    = -\frac{n_{m}u^{2}_{x}}{4}\begin{pmatrix}
 0 & 0 & b^{2} & 0 & 0\\
        0 & 2ab & 0 & 2b^{2} & 0 \\
        a^{2} & 0 & 4ab & 0 & b^{2} \\
        0 & 2a^{2} & 0 & 2ab & 0\\
       0 & 0 & a^{2} & 0 & 0
\end{pmatrix}\sum_{m=-2}^{2}\sum_{n=-2}^{2}e^{im\phi_{1}}e^{in\phi_{2}}
\end{align}
\begin{itemize}
    \item \textbf{y-component}
\end{itemize}
\begin{equation}
  U_{\mathbf{k_{1},k_{2}}}   = \frac{1}{2}\sum_{i}u^{i}_{y}e^{(\mathbf{k'-k}).\mathbf{R}_{i}}\left[a(e^{-i\phi_{2}}-e^{i\phi_{1}})+b(e^{-i\phi_{1}}-e^{i\phi_{2}})\right]i
\end{equation}
\begin{equation}
     U_{\mathbf{-k_{1},-k_{2}}}   = \frac{1}{2}\sum_{j}u^{j}_{y}e^{-(\mathbf{k'-k})\cdot\mathbf{R}_{j}}\left[a(e^{-i\phi_{2}}-e^{i\phi_{1}})+b(e^{-i\phi_{1}}-e^{i\phi_{2}})\right](-i)
\end{equation}
\begin{align}
    &\langle U_{\mathbf{k_1},\mathbf{k}_{2}}U_{\mathbf{-k_1},\mathbf{-k_2}}\bigl \rangle_{y}= \frac{n_{m}u^{2}_{y}}{4}\left[a(e^{-i\phi_{2}}-e^{i\phi_{1}})+b(e^{-i\phi_{1}}-e^{i\phi_{2}})\right]^{2}\nonumber\\&
    =\frac{n_{m}u^{2}_{y}}{4}[a^{2}e^{-2i\phi_{2}}+a^{2}e^{2i\phi_{1}}-2a^{2}e^{i\phi_{1}}e^{-i\phi_{2}}+b^{2}e^{2i\phi_{2}}+b^{2}e^{-2i\phi_{1}}-2b^{2}e^{-i\phi_{1}}e^{i\phi_{2}}+2abe^{-i\phi_{1}}e^{-i\phi_{2}}\nonumber\\&
    -4ab+2abe^{i\phi_{1}}e^{i\phi_{2}}],\nonumber \\&
    = \frac{n_{m}u^{2}_{y}}{4}\begin{pmatrix}
 0 & 0 & b^{2} & 0 & 0\\
        0 & 2ab & 0 & -2b^{2} & 0 \\
        a^{2} & 0 & -4ab & 0 & b^{2} \\
        0 & -2a^{2} & 0 & 2ab & 0\\
       0 & 0 & a^{2} & 0 & 0
\end{pmatrix}\sum_{m=-2}^{2}\sum_{n=-2}^{2}e^{im\phi_{1}}e^{in\phi_{2}}
\end{align}
\begin{itemize}
    \item \textbf{z-component}
    \begin{equation}
    U_{\mathbf{k_{1},k_{2}}}    = \frac{1}{2}\sum_{i}u^{i}_{z}e^{(\mathbf{k'-k}).\mathbf{R}_{i}}\left[a^{2}e^{i(\phi-\phi')}-b^{2}e^{-i(\phi-\phi')}\right]  
    \end{equation}
\end{itemize}
\begin{equation}
     U_{\mathbf{-k_{1},-k_{2}}}    = \frac{1}{2}\sum_{j}u^{j}_{z}e^{-(\mathbf{k'-k}).\mathbf{R}_{j}}\left[a^{2}e^{i(\phi-\phi')}-b^{2}e^{-i(\phi-\phi')}\right]  
    \end{equation}
\begin{align}
      &\langle U_{\mathbf{k_1},\mathbf{k}_{2}}U_{\mathbf{-k_1},\mathbf{-k_2}}\bigl \rangle_{z}=\frac{n_{m}u^{2}_{z}}{4}[a^{2}e^{i(\phi-\phi')}-b^{2}e^{-i(\phi-\phi')}]^{2},\nonumber\\&
      =\frac{n_{m}u^{2}_{z}}{4}[a^{4}e^{2i\phi_{1}}e^{-2i\phi_{2}}+b^{4}e^{-2i\phi_{1}}e^{2i\phi_{2}}-2a^{2}b^{2}], \nonumber\\&\frac{n_{m}u^{2}_{z}}{4}\begin{pmatrix}
      0 & 0 & 0 & 0 & b^{4}\\
        0 & 0 & 0 & 0 & 0 \\
        0 & 0 & -2a^{2}b^{2} & 0 & 0 \\
        0 & 0 & 0 & 0 & 0\\
       a^{4} & 0 & 0 & 0 & 0
\end{pmatrix}\sum_{m=-2}^{2}\sum_{n=-2}^{2}e^{im\phi_{1}}e^{in\phi_{2}},
      \end{align}
we have
\begin{equation}
  \frac{n_{m}u^{2}_{x}}{4}=\frac{\hbar}{2\pi N_{F}\tau}\frac{\tau}{2\tau_{m,x}(1+ab\cos2\phi)}  
\end{equation}
therefore
 \begin{equation}
     \langle U_{\mathbf{k_1},\mathbf{k}_{2}}U_{\mathbf{-k_1},\mathbf{-k_2}}\bigl \rangle_{x}=\frac{\hbar}{2\pi N_{F}\tau}\sum_{m=-2}^{2}\sum_{n=-2}^{2}z^{x}_{mn}e^{im\phi_{1}}e^{in\phi_{2}}
 \end{equation}

where we have $\alpha_{x}=\frac{\tau}{\tau_{m,x}}$, and 
\begin{equation}
z^{x}_{mn}= \begin{pmatrix}
 0 & 0 & -\frac{\alpha_{x}b^{2}}{2(1+ab\cos2\phi)} & 0 & 0\\
        0 & -\frac{\alpha_{x}ab}{(1+ab\cos2\phi)} & 0 & -\frac{\alpha_{x}b^{2}}{(1+ab\cos2\phi)} & 0 \\
        -\frac{\alpha_{x}a^{2}}{2(1+ab\cos2\phi)} & 0 & -\frac{\alpha_{x}2ab}{(1+ab\cos2\phi)} & 0 & -\frac{\alpha_{x}b^{2}}{2(1+ab\cos2\phi)} \\
        0 & -\frac{\alpha_{x}a^{2}}{(1+ab\cos2\phi)} & 0 & -\frac{\alpha_{x}ab}{(1+ab\cos2\phi)} & 0\\
       0 & 0 & -\frac{\alpha_{x}a^{2}}{2(1+ab\cos2\phi)} & 0 & 0
\end{pmatrix}. 
\end{equation}

\begin{equation}
  \frac{n_{m}u^{2}_{y}}{4}=\frac{\hbar}{2\pi N_{F}\tau}\frac{\tau}{2\tau_{m,y}(1-ab\cos2\phi)},  
\end{equation}
for the $y$ component
\begin{equation}
    \langle U_{\mathbf{k_1},\mathbf{k}_{2}}U_{\mathbf{-k_1},\mathbf{-k_2}}\bigl \rangle_{y}=\frac{\hbar}{2\pi N_{F}\tau}\sum_{m=-2}^{2}\sum_{n=-2}^{2}z^{y}_{mn}e^{im\phi_{1}}e^{in\phi_{2}},
\end{equation}
where we have $\alpha_{y}=\frac{\tau}{\tau_{m,y}}$, and 
\begin{equation}
z^{y}_{mn}= \begin{pmatrix}
 0 & 0 & \frac{\alpha_{y}b^{2}}{2(1-ab\cos2\phi)} & 0 & 0\\
        0 & \frac{\alpha_{y}ab}{(1-ab\cos2\phi)} & 0 & -\frac{\alpha_{y}b^{2}}{(1-ab\cos2\phi)} & 0 \\
        \frac{\alpha_{y}a^{2}}{2(1-ab\cos2\phi)} & 0 & -\frac{\alpha_{x}2ab}{(1-ab\cos2\phi)} & 0 & \frac{\alpha_{x}b^{2}}{2(1-ab\cos2\phi)} \\
        0 & -\frac{\alpha_{y}a^{2}}{(1-ab\cos2\phi)} & 0 & \frac{\alpha_{x}ab}{(1-ab\cos2\phi)} & 0\\
       0 & 0 & \frac{\alpha_{y}a^{2}}{2(1-ab\cos2\phi)} & 0 & 0
\end{pmatrix}. 
\end{equation}

   with
\begin{align}
      \frac{n_{m}u_{z}^{2}}{4}=  \frac{\hbar}{2\pi N_{F}\tau}\frac{\tau}{\tau_{z}(a^{4}+b^{4})} \\
    \alpha_{z}= \frac{\tau}{\tau_{m,z}}
\end{align}
and we have
\begin{equation}
    \langle U_{\mathbf{k_1},\mathbf{k}_{2}}U_{\mathbf{-k_1},\mathbf{-k_2}}\bigl \rangle_{z}=\frac{\hbar}{2\pi N_{F}\tau}\sum_{m=-2}^{2}\sum_{n=-2}^{2}z^{z}_{mn}e^{im\phi_{1}}e^{in\phi_{2}},
\end{equation}
where
\begin{equation}
      z^{z}_{mn}=  \begin{pmatrix}
     0 & 0 & 0 & 0 & \frac{\alpha_{z}b^{4}}{(a^{4}+b^{4})}\\
        0 & 0 & 0 & 0 & 0 \\
        0 & 0 & \frac{-2\alpha_{z}a^{2}b^{2}}{(a^{4}+b^{4})} & 0 & 0 \\
        0 & 0 & 0 & 0 & 0\\
       \frac{\alpha_{z}a^{4}}{(a^{4}+b^{4})} & 0 & 0 & 0 & 0
\end{pmatrix}.
\end{equation}
Finally, the bare vertex  is given by
\begin{equation}
\Gamma^{0}_{\mathbf{k_{1},k_{2}}}=\frac{\hbar}{2\pi N_{F}\tau}\sum_{m=-2}^{2}\sum_{n=-2}^{2}z_{mn}e^{im\phi_{1}}e^{in\phi_{2}},
\end{equation}
with
\begin{equation}
    \mathbf{z}=z_{mn} =z^{e}_{mn}+z^{x}_{mn}+z^{y}_{mn}+z^{z}_{mn}.
\end{equation}
If we have in-plane isotropy $(u_{x} = u_{y})$ then the bare vertex is given by
\begin{equation}
   \Gamma^{0}_{\mathbf{k_{1},k_{2}}}=\frac{\hbar}{2\pi N_{F}\tau}\sum_{m=-2}^{2}\sum_{n=-2}^{2}z^{iso}_{mn}e^{im\phi_{1}}e^{in\phi_{2}}, 
\end{equation}
where $z^{iso}_{mn}$ is given as
\begin{align}
     z^{iso}= 
    \begin{pmatrix}
    0 & 0 & 0 & 0 & z^{(-22)}\\
     0 & 0 & 0 & z^{(-11)} & 0 \\
      0 & 0 & z^{(00)} & 0 & 0 \\
      0 & z^{(1-1)} & 0 & 0 & 0 \\
      z^{(2-2)} & 0 & 0 & 0 & 0 &
    \end{pmatrix},
\end{align}
and the elements are given by 
\begin{align}
 z^{(-22)} &= \frac{b^4 \alpha _e}{a^4+b^4+1}+\frac{b^4 \alpha _z}{a^4+b^4}\nonumber\\
z^{(-11)} &= \frac{2 b^2 \alpha _e}{a^4+b^4+1}-2 b^2 \alpha _x \nonumber\\ 
z^{(00)} &= \frac{\left(2 a^2 b^2+1\right) \alpha _e}{a^4+b^4+1}-\frac{2 a^2 b^2 \alpha _z}{a^4+b^4}-4 a b \alpha _x \nonumber\\ 
z^{(1-1)} &= \frac{2 a^2 \alpha _e}{a^4+b^4+1}-2 a^2 \alpha _x \nonumber\\
z^{(2-2)} &= \frac{a^4 \alpha _e}{a^4+b^4+1}+\frac{a^4 \alpha _z}{a^4+b^4}. 
\end{align}
The Bethe-Salpeter equation for full vertex correction is given by,
\begin{figure}[h]
    \centering
    \includegraphics[width=0.5\textwidth]{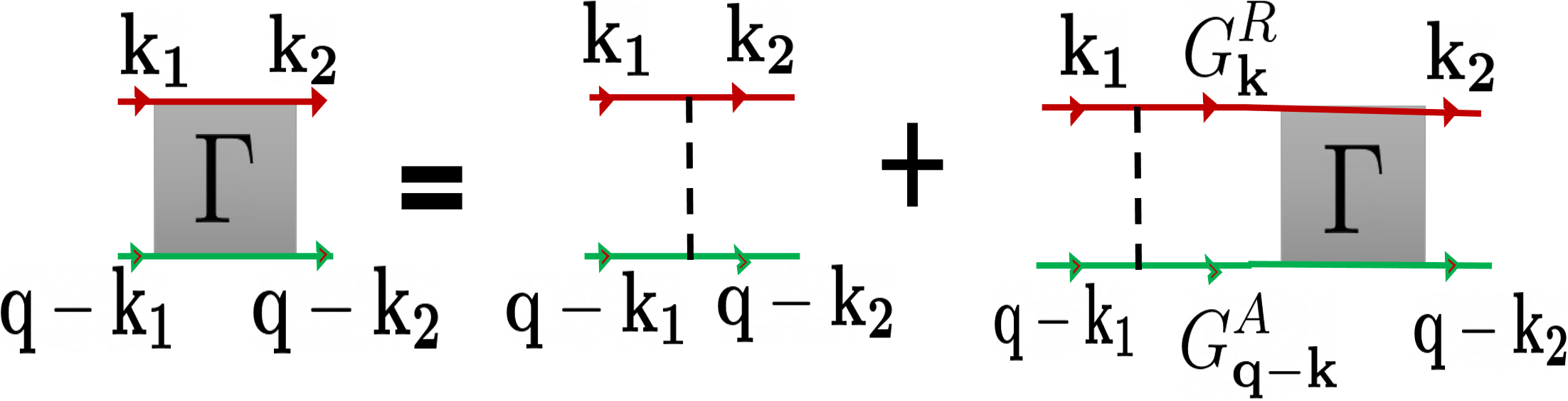}
    \caption{The Feynman diagram of the Bethe-Salpeter equation for the intravalley cooperons. The valley is conserved during scattering.}
    \end{figure}
\begin{equation}
    \Gamma_{\mathbf{k}_{1}, \mathbf{k}_{2}}= \Gamma_{\mathbf{k}_{1}, \mathbf{k}_{2}}^{0}+\sum_{\mathbf{k}}\Gamma_{\mathbf{k}_{1}, \mathbf{k}}^{0} G_{\mathbf{k}}^{R}G_{\mathbf{q}-\mathbf{k}}^{A} \Gamma_{\mathbf{k}, \mathbf{k}_{2}}.
\end{equation}
The above equation can be solved by expanding $  \Gamma_{\mathbf{k}_{1}, \mathbf{k}_{2}}$ as,
\begin{equation}
    \Gamma_{\mathbf{k_{1},k_{2}}}=\frac{\hbar}{2\pi N_{F}\tau}\sum_{m=-2}^{2}\sum_{n=-2}^{2}V^{mn}e^{im\phi_{1}}e^{in\phi_{2}},
\end{equation}
and the expansion coefficients $V^{mn}$ in matrix form can be obtained by solving
\begin{equation}
    \mathbf{V=(I-z\Phi)^{-1}z},
\end{equation}
where $\mathbf{z}$ is the $5\times5$ matrix for bare vertex, $\mathbf{I}$ is $5\times 5$ identity matrix, and $\Phi_{mn}$ is defined as
\begin{equation}
    \Phi_{mn}= \frac{1}{2\pi}\int_{0}^{2\pi}d\phi\frac{e^{i(m+n)\phi}}{1+i\tau\mathbf{q\cdot v_{F}}}.
\end{equation}
Up to $q^2$ terms in the small $\mathbf{q}$ limit 
\begin{equation}
  \mathbf{\Phi} \approx   \begin{pmatrix}
0 & 0 & -\frac{Q^{2}}{4} & -\frac{iQ}{2} & 1-\frac{Q^{2}}{2}\\
0 & -\frac{Q^{2}}{4} & -\frac{iQ}{2} &  1-\frac{Q^{2}}{2} &  -\frac{iQ}{2} \\
-\frac{Q^{2}}{4} &  -\frac{iQ}{2} &  1-\frac{Q^{2}}{2} & -\frac{iQ}{2} & -\frac{Q^{2}}{4}\\
-\frac{iQ}{2} & 1-\frac{Q^{2}}{2} & -\frac{iQ}{2} & -\frac{Q^{2}}{4} & 0 \\
1-\frac{Q^{2}}{2} & -\frac{iQ}{2} & -\frac{Q^{2}}{4} & 0 & 0
\end{pmatrix}
\end{equation}
where $Q=qv_{F}\tau$. We obtain the solution to the expansion coefficients with the assumptions of in-plane isotropy
\begin{equation}
     V= 2\begin{bmatrix}
        0 & 0 & -\frac{Q^{2}}{2} & -iQ & g^{(-22)}+ Q^{2}   \\
        0 & -\frac{Q^{2}}{2} & -iQ & g^{(-11)}+Q^{2} & -iQ   \\
      -\frac{Q^{2}}{2} & -iQ & g^{(00)}+Q^{2} & -iQ & -\frac{Q^{2}}{2} \\
      -iQ &  g^{(1-1)}+Q^{2} &-iQ &-\frac{Q^{2}}{2} & 0 \\
      g^{(2-2)}+Q^{2} & -iQ & -\frac{Q^{2}}{2} & 0 & 0
      \end{bmatrix}^{-1},
\end{equation}
where the ``Cooperon" gaps have been introduced:
\begin{align}
    g^{(-ii)} = 2(1-z^{(-ii)})/z^{(-ii)},
\end{align}
where $i$ runs from -2 to +2. 
The elements of the matrix $V$ are given by 
\scriptsize
\begin{align}
    {V^{-22}} &= 
    \frac{2}{Q^2 \left(\frac{{g^{(-22)}} {g^{(2-2)}} ({g^{(-11)}} ({g^{(00)}}+{g^{(1-1)}}+1)+{g^{(00)}} {g^{(1-1)}}+
    {g^{(1-1)}})+{g^{(-22)}} {g^{(-11)}} {g^{(00)}} ({g^{(1-1)}}+1)+({g^{(-11)}}+1) {g^{(00)}} {g^{(1-1)}}
   {g^{(2-2)}}}{g^{(-22)}g^{(-11)}g^{(00)}g^{(1-1)}}\right)+{g^{(2-2)}}} \nonumber\\
   V^{-11} &= \frac{2 }{Q^2 \left(\frac{{g^{(-22)}} {g^{(2-2)}} ({g^{(-11)}} ({g^{(00)}}+{g^{(1-1)}}+1)+{g^{(00)}} {g^{(1-1)}}+
    {g^{(1-1)}})+{g^{(-22)}} {g^{(-11)}} {g^{(00)}} ({g^{(1-1)}}+1)+({g^{(-11)}}+1) {g^{(00)}} {g^{(1-1)}}
   {g^{(2-2)}}}{g^{(-22)}g^{(-11)}g^{(00)}g^{(2-2)}}\right)+{g^{(1-1)}}} \nonumber\\
   V^{00} &= \frac{2 }{Q^2 \left(\frac{{g^{(-22)}} {g^{(2-2)}} ({g^{(-11)}} ({g^{(00)}}+{g^{(1-1)}}+1)+{g^{(00)}} {g^{(1-1)}}+
    {g^{(1-1)}})+{g^{(-22)}} {g^{(-11)}} {g^{(00)}} ({g^{(1-1)}}+1)+({g^{(-11)}}+1) {g^{(00)}} {g^{(1-1)}}
   {g^{(2-2)}}}{g^{(-22)}g^{(-11)}g^{(1-1)}g^{(2-2)}}\right)+{g^{(00)}}} \nonumber\\
   V^{1-1} &= \frac{2 }{Q^2 \left(\frac{{g^{(-22)}} {g^{(2-2)}} ({g^{(-11)}} ({g^{(00)}}+{g^{(1-1)}}+1)+{g^{(00)}} {g^{(1-1)}}+
    {g^{(1-1)}})+{g^{(-22)}} {g^{(-11)}} {g^{(00)}} ({g^{(1-1)}}+1)+({g^{(-11)}}+1) {g^{(00)}} {g^{(1-1)}}
   {g^{(2-2)}}}{g^{(-22)}g^{(00)}g^{(1-1)}g^{(2-2)}}\right)+{g^{(-11)}}} \nonumber\\
   V^{2-2} &= \frac{2 }{Q^2 \left(\frac{{g^{(-22)}} {g^{(2-2)}} ({g^{(-11)}} ({g^{(00)}}+{g^{(1-1)}}+1)+{g^{(00)}} {g^{(1-1)}}+
    {g^{(1-1)}})+{g^{(-22)}} {g^{(-11)}} {g^{(00)}} ({g^{(1-1)}}+1)+({g^{(-11)}}+1) {g^{(00)}} {g^{(1-1)}}
   {g^{(2-2)}}}{g^{(-11)}g^{(00)}g^{(1-1)}g^{(2-2)}}\right)+{g^{(-22)}}} \nonumber\\
   \label{Eq_V_general}
\end{align}
\normalsize
\section{Graphene}
First, we do the analysis for graphene ($a=1$, $b=0$). In this case $A^{(0)} = A^{(1)} = 0$, therefore $g^0=g^1\rightarrow\infty$. The elements of the matrix $V$ are given by 
\begin{align}
   V^{-22} &= \frac{2}{Q^2 \left(\frac{{g^{(00)}}
   ({g^{(1-1)}}+{g^{(2-2)}}+1)+{g^{(1-1)}} {g^{(2-2)}}+{g^{(2-2)}}}{g^{(00)}g^{(1-1)}}\right)+{g^{(2-2)}}} \nonumber\\
   V^{-11} &= \frac{2}{Q^2 \left(\frac{{g^{(00)}}
   ({g^{(1-1)}}+{g^{(2-2)}}+1)+{g^{(1-1)}} {g^{(2-2)}}+{g^{(2-2)}}}{g^{(00)}g^{(2-2)}}\right)+{g^{(1-1)}}} \nonumber\\
    V^{00} &= \frac{2}{Q^2 \left(\frac{{g^{(00)}}
   ({g^{(1-1)}}+{g^{(2-2)}}+1)+{g^{(1-1)}} {g^{(2-2)}}+{g^{(2-2)}}}{g^{(1-1)}g^{(2-2)}}\right)+{g^{(00)}}}\nonumber\\
   V^{1-1}&=0 \nonumber\\
   V^{2-2}&=0
\end{align}
In terms of the scattering time ratios ($\alpha_i=\tau\tau_i^{-1}$), the gaps are given by  
\begin{align}
    g^{(00)}&=-2+\frac{4}{\alpha_e} \nonumber\\
    g^{(1-1)}&=-2+\frac{2}{\alpha_e-2\alpha_x} \nonumber\\
    g^{(2-2)}&=-2+\frac{4}{\alpha_e+2\alpha_z}
\end{align}
Let us also recall the following relation: 
\begin{align}
    \tau^{-1} &= \tau_e^{-1} + \tau_z^{-1} + 2\tau_x^{-1}\nonumber\\
    1&=(\tau_e^{-1} + \tau_z^{-1} + 2\tau_x^{-1})\tau\nonumber\\
    1&=\alpha_e+\alpha_z+2\alpha_x
    \label{eq_alpha_sum}
\end{align}
If we want the elements in the matrix $V$ to diverge, then the gaps must vanish. 

Now $g^{(00)}=0$ implies that $\alpha_e=2$, which is not possible due to the constraint in Eq.~\ref{eq_alpha_sum}. Therefore $g^{(00)}=0$ is never satisfied. 

The constraint $g^{(1-1)}=0$ implies that $\alpha_e-2\alpha_x=1$. This, along with Eq.~\ref{eq_alpha_sum}, implies that $\alpha_e=1$ and $\alpha_z=\alpha_x=0$ is the only solution. Therefore $V^{-11}$ can diverge, which can result in weak antilocalization. 

The constraint $g^{(2-2)}=0$ implies that $\alpha_e+2\alpha_z=2$. This along with Eq.~\ref{eq_alpha_sum} implies that $\alpha_e=4\alpha_x$, and $\alpha_z=1-3\alpha_e/2$. This term results in weak localization, which is caused by magnetic impurities. 

For Graphene, when $\alpha_e=1$, we recover $\eta_\nu=2$, and $\eta_H=-1/4$. The vertex is 
\begin{equation}
    \Gamma_{\mathbf{q}}=\frac{\hbar}{2\pi N_{F}\tau}\left(V^{-22}-V^{-11}+V^{00}\right),
\end{equation}
We retain the most divergent terms in the vertex, and thus 
\begin{align}
    \Gamma_{\mathbf{q}}=\frac{\hbar}{2\pi N_{F}\tau}\left(\frac{2}{g^{(2-2)}+Q^2\left(1+\frac{1}{g^{(1-1)}}\right)}-\frac{2}{g^{(1-1)}+Q^2\left(1+\frac{1}{g^{(00)}}+\frac{1}{g^{(2-2)}}\right)}\right),
\end{align}
and $Q=qv_{F}\tau$. 
Note that $V^{00}$ is not retained as $g^{(00)}$ can never vanish in the limit $q\rightarrow 0$. 
\subsection{Quantum interference correction to conductivity
at zero field $\sigma^{F}(0)$ for Graphene}
The zero-field quantum interference correction to conductivity is

\begin{align}
     \sigma^{F}(0)= -\frac{e^{2}}{\pi h}\sum_{i= 0}^{1}\int_{\ell^{-2}_{\phi}}^{\ell^{-2}_{e}}d(q^{2})\frac{\alpha_{i}}{\ell^{-2}_{i}+q^{2}}, \\
    = -\frac{e^{2}}{\pi h}\sum_{i= 0}^{1}\alpha_{i} \ln\frac{\ell^{-2}_{i}+\ell^{-2}_{e}}{\ell^{-2}_{i}+\ell^{-2}_{\phi}}
\end{align}
with
\begin{align}
      \alpha_{0}= \frac{\eta^{2}_{v}(1+2\eta_{H})}{2(1+\frac{1}{g^{(1-1)}})}, \quad {\ell_{0}}^{-2} = \frac{g^{(2-2)}}{2\ell^{2}(1+\frac{1}{g^{(1-1)}})} 
\end{align}
\begin{align}
     \alpha_{1}= -\frac{\eta^{2}_{v}(1+2\eta_{H})}{2(\frac{1}{g^{(2-2)}}+\frac{1}{g^{(00)}}+1)}, \quad \ell^{-2}_{1} = \frac{g^{(1-1)}}{2\ell^{2}(\frac{1}{g^{(2-2)}}+\frac{1}{g^{(00)}}+1)},
\end{align}
and $\ell=\sqrt{D\tau}$ with $D=\frac{v^{2}_{F}\tau}{2}$ is the diffusion constant. $\tau_{e}$ and  $\tau_{m}$ are related to the elastic scattering length $\ell_{e}$ and magnetic scattering length $\ell_{m}$ by $\ell_{e}=\sqrt{D\tau_{e}}$ and $\ell_{m}=\sqrt{D\tau_{m}}$. The quantum diffusion condition is reflected in the upper bound ${q_{max}=\ell^{-1}_e}$ and the lower bound ${q_{min}=\ell^{-1}_\phi}$ of the integration over \textit{q} and $1/\ell^{2} \equiv 1/\ell^{2}_{e}+1/\ell^{2}_{m}$. We assume that $\ell_{e}$ is much shorter than phase coherence length $\ell_{\phi}$, as required by the quantum diffusive transport. The first term of Eq. (135) gives \textbf{weak localization}, and the second term is for \textbf{weak anti-localization}.
\subsection{Magnetoconductivity for Graphene}
In the presence of a perpendicular magnetic field $\textit{B}$, the finite-field conductivity correction $\sigma^{F}(B)$ can be derived by replacing $q^2$ in Eq.(133) with $q^{2}_{n}=(n+\frac{1}{2})/\ell^{2}_{B}$, where $n$ labels the Landau levels for 2D Dirac fermions and $\ell_{B}=\sqrt{\hbar/4eB}$ is defined as the magnetic length for a Cooperon. Therefore the finite-field conductivity correction becomes
\begin{align}
    \sigma^{F}(B)=-\frac{e^{2}}{\pi h}\sum_{i=0, 1}\alpha_{i}\left[\Psi\left(\frac{\ell^{2}_{B}}{\ell^{2}_{e}}+\frac{\ell^{2}_{B}}{\ell^{2}_{i}}+\frac{1}{2}\right)-\Psi\left(\frac{\ell^{2}_{B}}{\ell^{2}_{\phi}}+\frac{\ell^{2}_{B}}{\ell^{2}_{i}}+\frac{1}{2}\right)\right]
\end{align}
where $\Psi$ is the digamma function. The magnetoconductivity is found as
\begin{multline}
    \Delta\sigma(B) \equiv  \sigma^{F}(B)- \sigma^{F}(0) \\ 
    =\frac{e^{2}}{\pi h}\sum_{i=0, 1}\alpha_{i}\left[\Psi\left(\frac{\ell^{2}_{B}}{\ell^{2}_{\phi}}+\frac{\ell^{2}_{B}}{\ell^{2}_{i}}+\frac{1}{2}\right)-\ln\left(\frac{\ell^{2}_{B}}{\ell^{2}_{\phi}}+\frac{\ell^{2}_{B}}{\ell^{2}_{i}}\right)\right],
\end{multline}
where we have assumed a small magnetic field limit, in which $\ell_{B}\gg \ell_{e}$ and 
\begin{align}
    \Psi\left(\frac{\ell^{2}_{B}}{\ell^{2}_{e}}+\frac{\ell^{2}_{B}}{\ell^{2}_{i}}+\frac{1}{2}\right) \approx \ln\left(\frac{\ell^{2}_{B}}{\ell^{2}_{e}}+\frac{\ell^{2}_{B}}{\ell^{2}_{i}}\right).
\end{align}

\section{Dice lattice}

The Cooperon gaps are 
\begin{align}
    g^{(-22)}&=\frac{12}{\alpha _e+3 \alpha _z}-2\nonumber\\
    g^{(-11)}&=\frac{6}{2 \alpha _e-3 \alpha _x}-2\nonumber\\
    g^{(00)}&=\frac{2}{\alpha _e-2 \alpha _x-\alpha _z}-2\nonumber\\
    g^{(1-1)}&=\frac{6}{2 \alpha _e-3 \alpha _x}-2\nonumber\\
    g^{(2-2)}&=\frac{12}{\alpha _e+3 \alpha _z}-2
\end{align}

In the limit when $q\rightarrow 0$, the relevant elements of $V$ will diverge if the corresponding Cooperon gaps vanish. So we will analyze each case. 

$g^{(-22)}=0$ implies that $\alpha_e+3\alpha_z=6$. Since $\alpha_e\leq 1$ and $\alpha_z\leq 1$, this condition is never satisfied. 

$g^{(-11)}=0$ implies that $2\alpha_e-3\alpha_z=3$. This condition is also never satisfied for the same reason as above. 

$g^{(00)}=0$ implies that $\alpha_e-2\alpha_x-\alpha_z = 1$. This condition is satisfied when $\alpha_e=1$, $\alpha_x=\alpha_z=0$. Hence $V^{00}$ can diverge, yielding weak localization. 

The conditions $g^{(1-1)}=0$ and $g^{(2-2)}=0$ are the same as conditions 1 and 2. 
For the Dice lattice, we have 
\begin{align}
    \eta_\nu = \frac{1}{1-\left(\alpha_x+\frac{2\alpha_e}{3}\right)}\nonumber\\
    \eta_H = -\frac{1}{2}\left(1-\frac{1}{\eta_\nu}\right)
\end{align}
Again, retaining the most divergent terms in the lattice, we have 
\begin{align}
    \Gamma_{\mathbf{q}}=\frac{\hbar}{2\pi N_{F}\tau}\left(\frac{2}{g^{(00)}+Q^2\left(1+\frac{1}{g^{(-11)}}\right)}\right)
\end{align}
\subsection{Quantum interference correction to the conductivity
at zero field $\sigma^{F}(0)$ for Dice lattice}
Quantum interference correction at zero-field for the Dice lattice is found to be
\begin{align}
     \sigma^{F}(0)= -\frac{e^{2}}{\pi h}\int_{\ell^{-2}_{\phi}}^{\ell^{-2}_{e}}d(q^{2})\frac{\beta_{0}}{\ell^{-2}_{0}+q^{2}} \\ =
     -\frac{e^{2}}{\pi h}\beta_{0} \ln\frac{\ell^{-2}_{0}+\ell^{-2}_{e}}{\ell^{-2}_{0}+\ell^{-2}_{\phi}}
\end{align}
\begin{align}
  \beta_{0}= \frac{\eta^{2}_{v}(1+2\eta_{H})}{2(1+\frac{1}{g^{(-11)}})}, \quad \ell^{-2}_{0} = \frac{g^{(00)}}{2\ell^{2}(1+\frac{1}{g^{(-11)}})},    
\end{align}
this correction gives rise to the phenomena of \textbf{weak-localization}.
\subsection{Magnetoconductivity for the Dice lattice}

The magnetoconductivity is found to be
\begin{align}
    &\Delta\sigma(B) \equiv  \sigma^{F}(B)- \sigma^{F}(0) \nonumber\\ 
    &=\frac{e^{2}}{\pi h}\beta_{0}\left[\Psi\left(\frac{\ell^{2}_{B}}{\ell^{2}_{\phi}}+\frac{\ell^{2}_{B}}{\ell^{2}_{i}}+\frac{1}{2}\right)-\ln\left(\frac{\ell^{2}_{B}}{\ell^{2}_{\phi}}+\frac{\ell^{2}_{B}}{\ell^{2}_{i}}\right)\right].
\end{align}
\section{Elastic impurities} 

We now examine the case when $\alpha_e=1$, $\alpha_x=\alpha_z=0$, and look for the possible solutions of $a$ and $b$. Since $a^2+b^2=1$, we can now express the Cooperon gaps solely in terms of $a$ or $b$. The Cooperon gaps are given by 

\begin{align}
    g^{(-22)}&=\frac{4}{b^4}-\frac{4}{b^2}+2\nonumber\\
    g^{(-11)}&=2 \left(b^2+\frac{1}{b^2}-2\right)\nonumber\\
    g^{(00)}&=\frac{6}{-2 b^4+2 b^2+1}-4\nonumber\\
    g^{(1-1)}&=-\frac{2 b^4}{b^2-1}\nonumber\\
    g^{(2-2)}&=\frac{2\left(b^4+1\right)}{\left(b^2-1\right)^2}
\end{align}

The only possible solutions when the Cooperon gaps vanish are: \\

$g^{(-11)}=0$, $b=1$, $a=0$: corresponding to WAL in Graphene.\\

$g^{(00)}=0$, $b=1/\sqrt{2}$, $a=1/\sqrt{2}$: corresponding to WL in Dice lattice.\\

$g^{(1-1)}=0$, $b=0$, $a=1$: corresponding to WAL in Graphene.

For the case of only elastic impurities, we have 
\begin{align}
    \eta_\nu = \frac{2b^4 - 2b^2 +2}{2b^4-2b^2+1}
\end{align}
We retain the most divergent terms in the vertex, and thus 
\begin{align}
    \Gamma_{\mathbf{q}}=\frac{\hbar}{2\pi N_{F}\tau}\left(-\frac{2}{g^{(1-1)}+Q^2\left(1+\frac{1}{g^{(00)}}\right) }+ \frac{2}{g^{(00)}+Q^2\left(1+\frac{1}{g^{(1-1)}}+\frac{1}{g^{(-11)}}\right)} - \frac{2}{g^{(-11)}+Q^2\left(1+\frac{1}{g^{(00)}}\right)} \right).
\end{align}
We have to collect the most divergent terms of vertex correction for which cooperon gaps must vanish in the limit of $\mathbf{q}\to 0$. 
\subsection{Quantum interference correction to the conductivity
at zero field $\sigma^{F}(0)$ for elastic impurities only}
For the case of only elastic impurities, we have

\begin{align}
     \sigma^{F}(0)= -\frac{e^{2}}{\pi h}\sum_{i= 0}^{2}\int_{\ell^{-2}_{\phi}}^{\ell^{-2}_{e}}d(q^{2})\frac{\beta_{i}}{\ell^{-2}_{i}+q^{2}}, \\
    = -\frac{e^{2}}{\pi h}\sum_{i= 0}^{2
    }\beta_{i} \ln\frac{\ell^{-2}_{i}+\ell^{-2}_{e}}{\ell^{-2}_{i}+\ell^{-2}_{\phi}}
\end{align}
\begin{align}
      \beta_{0}= -\frac{\eta^{2}_{v}(1+2\eta_{H})}{2(1+\frac{1}{g^{(00)}})}, \quad \ell^{-2}_{0} = \frac{g^{(1-1)}}{2l^{2}(1+\frac{1}{g^{(00)}})} 
\end{align}
\begin{align}
     \beta_{1}= \frac{\eta^{2}_{v}(1+2\eta_{H})}{2(\frac{1}{g^{(1-1)}}+\frac{1}{g^{(-11)}}+1)}, \quad \ell^{-2}_{1} = \frac{g^{(00)}}{2l^{2}(\frac{1}{g^{(1-1)}}+\frac{1}{g^{(-11)}}+1)},
\end{align}
\begin{align}
      \beta_{2}= -\frac{\eta^{2}_{v}(1+2\eta_{H})}{2(1+\frac{1}{g^{(00)}})}, \quad \ell^{-2}_{2} = \frac{g^{(-11)}}{2l^{2}(1+\frac{1}{g^{(00)}})} 
\end{align}
\subsection{Magnetocondcutivity in the presence of elastic impurities only}
The magnetoconductivity in the presence of elastic impurities can be found to be 
\begin{align}
    &\Delta\sigma(B) \equiv  \sigma^{F}(B)- \sigma^{F}(0) \nonumber\\ 
    &=\frac{e^{2}}{\pi h}\sum_{i=0,1, 2}\beta_{i}\left[\Psi\left(\frac{\ell^{2}_{B}}{\ell^{2}_{\phi}}+\frac{\ell^{2}_{B}}{\ell^{2}_{i}}+\frac{1}{2}\right)-\ln\left(\frac{\ell^{2}_{B}}{\ell^{2}_{\phi}}+\frac{\ell^{2}_{B}}{\ell^{2}_{i}}\right)\right],
\end{align}
\section{Magnetic impurities in the $z$-direction}
We now examine the case when $\alpha_z=1$, $\alpha_x=\alpha_e=0$, and look for the possible solutions of $a$ and $b$. Since $a^2+b^2=1$, we can now express the Cooperon gaps solely in terms of $a$ or $b$. The Cooperon gaps are given by 

\begin{align}
    g^{(-22)}&=\frac{2 \left(b^2-1\right)^2}{b^4}\nonumber\\
    g^{(-11)}&=\infty\nonumber\\
    g^{(00)}&=\frac{1}{b^2 \left(b^2-1\right)}\nonumber\\
    g^{(1-1)}&=\infty\nonumber\\
    g^{(2-2)}&=\frac{2 b^4}{\left(b^2-1\right)^2}
\end{align}
The only possible solutions are $g^{(-22)}=0$, $b=1$, $a=0$, and $g^{(2-2)}=0$, $a=1$, $b=0$, both corresponding to WL in graphene. We retain the most divergent terms in the vertex, and thus 
\begin{align}
    \Gamma_{\mathbf{q}}=\frac{\hbar}{2\pi N_{F}\tau}\left(\frac{2}{g^{(2-2)}+Q^2 }+ \frac{2}{g^{(-22)}+Q^2}\right),
\end{align}
where we have $\eta_{H}=0$ and $\eta_{v}=1$.
\subsection{Quantum interference correction to the conductivity
at zero field $\sigma^{F}(0)$ for the magnetic impurities in the z-direction }
For the case of magnetic impurities in the z-direction, we have
\begin{align}
     \sigma^{F}(0)= -\frac{e^{2}}{\pi h}\sum_{i= 0,1}\int_{\ell^{-2}_{\phi}}^{\ell^{-2}_{e}}d(q^{2})\frac{\gamma_{i}}{\ell^{-2}_{i}+q^{2}}, \\
    = -\frac{e^{2}}{\pi h}\sum_{i=0,1}\gamma_{i} \ln\frac{\ell^{-2}_{i}+\ell^{-2}_{e}}{\ell^{-2}_{i}+\ell^{-2}_{\phi}}
\end{align}
where we have
\begin{align}
    \gamma_{0}=\frac{1}{2}, \ \ell_{0}^{-2}=\frac{g^{(2-2)}}{2\ell^{2}} \\
    \gamma_{1}=\frac{1}{2}, \ell_{1}^{-2}=\frac{g^{(-22)}}{2\ell^{2}} 
\end{align}
This results in weak localization.
\subsection{Magnetocondcutivity in the presence of magnetic impurities in the z-direction}
The magnetoconductivity in the presence of magnetic impurities in the z-direction is found to be
\begin{align}
    &\Delta\sigma(B) \equiv  \sigma^{F}(B)- \sigma^{F}(0) \nonumber\\ 
    &=\frac{e^{2}}{\pi h}\sum_{i=0,1}\gamma_{i}\left[\Psi\left(\frac{\ell^{2}_{B}}{\ell^{2}_{\phi}}+\frac{\ell^{2}_{B}}{\ell^{2}_{i}}+\frac{1}{2}\right)-\ln\left(\frac{\ell^{2}_{B}}{\ell^{2}_{\phi}}+\frac{\ell^{2}_{B}}{\ell^{2}_{i}}\right)\right],
\end{align}
\section{Magnetic impurities in the $x$-direction}
We now examine the case when $\alpha_x=1$, $\alpha_z=\alpha_e=0$, and look for the possible solutions of $a$ and $b$. Since $a^2+b^2=1$, we can now express the Cooperon gaps solely in terms of $a$ or $b$. The Cooperon gaps are given by 

\begin{align}
    g^{(-22)}&=\infty\nonumber\\
    g^{(-11)}&=-\frac{1}{b^2}-2\nonumber\\
    g^{(00)}&=-\frac{1}{2 b \sqrt{1-b^2}}-2\nonumber\\
    g^{(1-1)}&=\frac{1}{b^2-1}-2\nonumber\\
    g^{(2-2)}&=\infty
\end{align}
None of the vanishing Cooperon gaps have any valid solution for $a$ and $b$. 
\end{widetext}
\end{document}